\documentclass[aps,pra,reprint,twocolumn,floatfix,notitlepage, floatfix,superscriptaddress,longbibliography]{revtex4-1}
\usepackage[utf8]{inputenc}
\usepackage{amsfonts,amsmath,amssymb,graphicx,epstopdf,verbatim}
\usepackage{bbold}
\usepackage{placeins}
\usepackage{rotating}
\usepackage{epstopdf}
\usepackage{xcolor}
\usepackage[colorlinks]{hyperref}
\definecolor{darkred}{rgb}{0.5,0,0}
\definecolor{darkgreen}{rgb}{0,0.5,0}
\definecolor{darkblue}{rgb}{0,0,0.5}
\hypersetup{colorlinks,
linkcolor=darkred,
filecolor=darkgreen,
urlcolor=darkblue,
citecolor=darkgreen}
\usepackage{braket}

\newcommand{\LL}{\mathcal{L}}

\newcommand{\rhot}{\hat{\rho}(t)}

\renewcommand{\Re}[1]{\mathbb{R}\mathrm{e}\left[#1\right]}

\newcommand{\sss}{\hat{\rho}_{\rm ss}}
\newcommand{\eig}[1]{\hat{\rho}_{#1}}

\newcommand{\Tr}[1]{\mathrm{Tr}\!\left[#1\right]}

\begin{document}

\title{Mean-field validity in a dissipative critical system: Liouvillian gap, \\ $\mathbb{PT}$-symmetric antigap, and permutational symmetry in the \emph{XYZ} model
}
\author{Dolf Huybrechts}\email{dolf.huybrechts@uantwerpen.be}
\affiliation{Theory of Quantum \& Complex Systems, University of Antwerp, B-2610 Wilrijk, Belgium}
\author{Fabrizio Minganti}\email{fabrizio.minganti@riken.jp}
\affiliation{Theoretical Quantum Physics Laboratory, RIKEN Cluster for Pioneering Research, Wako-shi, Saitama 351-0198, Japan}
\author{Franco Nori}\email{fnori@riken.jp}
\affiliation{Theoretical Quantum Physics Laboratory, RIKEN Cluster for Pioneering Research, Wako-shi, Saitama 351-0198, Japan}
\affiliation{Physics Department, The University of Michigan, Ann Arbor, Michigan, 48109-1040, USA}
\author{Michiel Wouters}\email{michiel.wouters@uantwerpen.be}
\affiliation{Theory of Quantum \& Complex Systems, University of Antwerp, B-2610 Wilrijk, Belgium}
\author{Nathan Shammah}\email{nathan.shammah@gmail.com}
\affiliation{Theoretical Quantum Physics Laboratory, RIKEN Cluster for Pioneering Research, Wako-shi, Saitama 351-0198, Japan}

\begin{abstract}
We study the all-to-all connected \emph{XYZ} (anisotropic-Heisenberg) spin model with \emph{local and collective} dissipations, comparing the results of mean field theory with the solution of the Lindblad quantum evolution.
Leveraging the permutational symmetry of the model [N. Shammah \emph{et al.}, \href{https://doi.org/10.1103/PhysRevA.98.063815}{Phys Rev. A {\bf 98}, 063815 (2018)}], we find exactly (up to numerical precision) the steady state up to $N=95$ spins.
We characterize criticality, studying, as a function of the number of spins $N$, the spin structure factor, the magnetization, the Liouvillian gap and the Von Neumann entropy of the steady state.
Exploiting the weak $\mathcal{PT}$-symmetry of the model, we efficiently calculate the Liouvillian gap, introducing the idea of an \emph{antigap}.
For small anisotropy, we find a paramagnetic-to-ferromagnetic phase transition in agreement with the mean-field theory.
For large anisotropy, instead, we find a significant discrepancy from the scaling of the low-anisotropy ferromagnetic phase.
We also study other more experimentally-accessible witnesses of the transition, which can be used for finite-size studies, namely the bimodality coefficient and the angular averaged susceptibility.
In contrast to the bimodality coefficient, the angular averaged susceptibility fails to capture the onset of the transition, in striking difference with respect to lower-dimensional studies.
We also analyze the competition between local dissipative processes (which disentangle the spin system) and collective dissipative ones (generating entanglement). 
The nature of the phase transition is almost unaffected by the presence of these terms.
Our results mark a stark difference with the common intuition that an all-to-all connected system should fall onto the mean-field solution also for intermediate number of spins.
\end{abstract}

\date{\today}

\maketitle

\tableofcontents

\section{Introduction}
\label{sec:intro}
Many-body quantum physics with light and matter is at the center of intense research, being at the crossroad of condensed matter, statistical mechanics, quantum optics, and quantum information.
In these open quantum systems, excitations, energy, and coherence are continuously exchanged with the environment, and they can be driven via pumping mechanisms \cite{Haroche_BOOK_Quantum,BreuerBookOpen,Carmichael_BOOK_1}.
Experimentally, light-matter interactions can be studied using Rydberg atoms confined between high-quality mirrors \cite{Haroche_BOOK_Quantum}, superconducting circuits \cite{SchoelkopfNat08,YouNat11}, semiconductor cavities \cite{DeveaudBOOK,Kavokin,BallariniNano19}, and optomechanical systems \cite{AspelmeyerRMP14}.
In many of these setups, a key role is played by the “photons”, that is, electromagnetic field excitations dressed by the matter degrees of freedom, thus permitting a finite effective photon-photon interaction (e.g., the polariton \cite{HopfieldPR58,CiutiSST03,Carusotto_RMP_2013_quantum_fluids_light}).

The experimental advances of the last decade provided the opportunity to realize extended lattices of resonators, allowing to explore criticality in this out-of-equilibrium context.
While quantum or thermal phase transitions can be determined by (free-)energy analysis \cite{Landau_BOOK_Statistical, SachdevBOOKPhase}, their dissipative counterparts need not to obey the same paradigm \cite{KesslerPRA12,TorrePRB2012, MarinoPRL2016,MingantiPRA18_Spectral,Kirton17,Kirton19rev}, and by properly designing the coupling with the environment and the driving mechanisms, it is possible to stabilize phases without an equilibrium counterpart \cite{DiehlNATPH2008,VerstraeteNATPH2009,Lambert09,Diehl10,LeePRL13,IlesSmithPRA14,JinPRX16}.
There exists a plethora of theoretical examples discussing the emergence of such dissipative phase transitions for photonic systems \cite{CarmichaelPRX15,WeimerPRL2015,BenitoPRA16,MendozaPRA16,CasteelsPRA16,BartoloPRA16,CasteelsPRA17,CasteelsPRA17-2,Foss-FeigPRA17,BiondiPRA17,BiellaPRA17,SavonaPRA17,Munuz2018,VicentiniPRA18,VerstraelenAS18}, lossy polariton condensates \cite{SiebererPRL13,SiebererPRB14,AltmanPRX15}, and spin models \cite{MorrisonPRA08,MorrisonPRL08,LeePRA11,KesslerPRA12,LeePRL13,LeePRA14,ChanPRA2015,JinPRX16,MaghrebiPRB16,RotaPRB17,OverbeckPRA17,RoscherarXiv18,RotaNJP18}.
Moreover, some key experiments proved the validity of the theoretical predictions in single superconducting cavities \cite{FinkPRX17} and lattices of superconducting resonators \cite{HouckNatPhys12,FitzpatrickPRX17}, Rydberg atoms in optical lattices \cite{Mueller_2012,BernienNAT2017}, optomechanical systems \cite{AspelmeyerRMP14,GilSantosPRL17}, exciton-polariton condensates \cite{KasprzakNAT2006,Carusotto_RMP_2013_quantum_fluids_light}, and semiconductor micropillars \cite{RodriguezPRL17,FinkNatPhys18}.

\begin{center}
\begin{figure}[!h]
  \centering
    \includegraphics[width=0.45\textwidth]{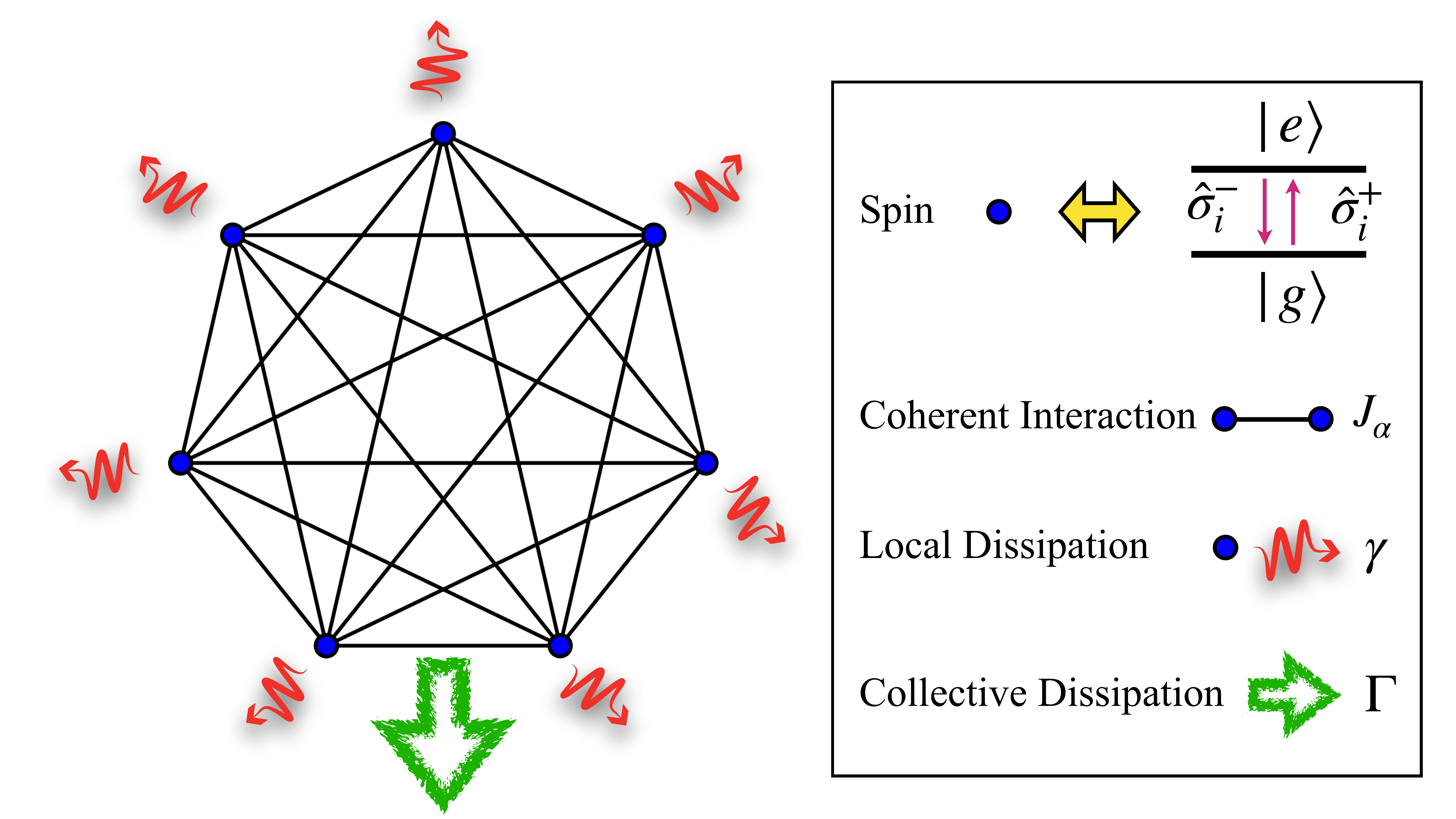}
      \caption{
      Sketch of the dissipative \emph{XYZ} model, with local and collective dissipation. In the legend we illustrate the possibility of implementing the spin model on an ensemble of two-level systems, or (artificial) atoms, interacting with an electromagnetic field. Each two-level system can switch between a ground, $|g\rangle$, and excited state, $|e\rangle$. While the spin-spin interactions, $\propto J_\alpha$, of the all-to-all connected lattice, can be mediated by the coherent interaction with the photonic field, its collective mode dissipates, at a rate $\propto\Gamma$, and all other spin-flip mechanisms contribute to local dissipation, $\propto\gamma$.}
\label{fig:xyz}
\end{figure}
\end{center}

In particular, the competition between interaction, driving and dissipation processes can lead to exotic physics, such as a transition from a photonic Mott insulator to a superfluid phase \cite{GreentreeNatPhys06,HartmannNatPhys06,AngelakisPRA07,HartmannLPR08,LebreuillyPRA17}, similar to that observed with ultracold atoms confined in optical lattices \cite{GreinerNature2002,BlochRMP08}.
Moreover, in the limit of a very strong nonlinearity one enters the regime of \textit{photon-blockade} \cite{CarmichaelPRL85,ImamogluPRL97,MiranowiczPRA13,KowalewskaPRA19}, where the presence of two photons inside the cavity becomes practically impossible.
This effect has been observed experimentally both in a single atom in a cavity \cite{BirnbaumNat05} and in a single superconducting circuit \cite{LangPRL11}. 
Interestingly, a system of coupled superconducting resonators \cite{AngelakisPRA07,HartmannPRL07,KayEPL08,HouckNatPhys12,PuriNatComm17} or Rydberg atoms \cite{LeePRA11,QianPRA12,ViteauPRL12,GlaetzlePRL15,QianPRA15,NguyenPRX18} can be mapped onto an effective spin model, as sketched in Fig.~\ref{fig:xyz}.

In this regard, the \emph{XYZ} Heisenberg model describes, with a high degree of generality, these systems and other spin models.
In the dissipative $XYZ$ model, each spin interacts with its nearest neighbors via an anisotropic Heisenberg Hamiltonian.
Moreover, each spin is coupled to the environment inducing random spin-flips in the $z$-axis direction. 
Due to its relative generality and simplicity, this model has been taken both as an example of a system exhibiting dissipative phase transitions, as well as a benchmark to test numerical methods.
Indeed, a single-site Gutzwiller mean-field (MF) theory can already retrieve a rich phase diagram for this model \cite{LeePRL13}. 
Numerical studies, capable of including long-range correlations, have confirmed a critical behavior in two-dimensional lattices and the absence of criticality in 1D \cite{JinPRX16,RotaPRB17,OrusNatComm17,BiellaPRA17,CasteelsPRA18,RotaNJP18, HuybrechtsPRA19}. 
Notwithstanding the fact that a collective bosonic field can be mapped onto an all-to-all-connected spin system \cite{WilsonPRA16}, we emphasize that the rich \emph{XYZ} model phase diagram in different regimes is a cornerstone of the study of many-body spin quantum systems, magnetism, spin dynamics and quantum phase transitions \cite{NguyenPRX18}. Indeed, it is the most general case of the Ising model and of the $XXZ$ model, of the Lipkin-Meshkov-Glick model and other spin-squeezing Hamiltonians, to which it can fall onto, for the appropriate choice of parameters \cite{LeePRA14}.

\subsection{This work}
\label{sec:thiswork}
In this article, we investigate the properties of an all-to-all (or fully) connected dissipative \emph{XYZ} system.
The interest and purpose of this study is manifold: 

(i) In the general study of a quantum system, one can think of the all-to-all connected model with uniform coupling as one in which long-range correlations cannot take place since all sites are at distance one. In this regard, it is ``common wisdom'' that a high-dimensional large system should recover the results of the mean-field prediction. Even if this can be argued for thermodynamic systems (where Landau-Ginzburg  theory can be applied to determine phase transitions \cite{PathriaBOOK}), the lack of free energy analysis does not allow such an easy argument in open quantum systems. We will consider the simplest type of non-thermal bath to try to address this question. 

(ii) Even if the mean field were to work, it should be predictive only in the thermodynamic limit. What is not clear is how the system behavior scales up to the infinite spin number. The high degree of symmetry of the all-to-all connected system allows for a dramatic reduction of the computational cost of the numerical calculations \cite{ShammahPRA98}. Moreover, many atoms-in-cavity experiments can be recast as all-to-all connected models by the mediation of the electromagnetic field, which collectively interacts with the atoms \cite{WilsonPRA16}. However, since in these systems there is a limited number of particles, identifying the correct observables to characterize the emergence of the phase transition is of paramount importance. We provide a thorough study of the spin structure factor, the collective magnetization, the bimodality coefficient and the angular averaged susceptibility. We also characterize less experimentally accesible quantities signaling the phase transition, as the Von Neumann entropy of the steady state and the Liouvillian spectrum and its gap. We test which one fares better in this intermediate regime to capture the onset of criticality.

(iii) The permutational method which we use here is exact (that is, no approximation on the model has been done). Exact computations on open-spin systems have been carried out for systems up to 16 spins \cite{RotaNJP18}. This article pushes this boundary far beyond this limit. 

The all-to-all connected geometry under consideration constitutes also an ideal benchmark for linked-cluster expansion theories \cite{NiuPRB89}.
In this kind of approach, one develops a perturbation expansion in power series of the coordination number around the Gutzwiller (or atomic) MF limit of a lattice model \cite{MetznerPRB91}. In the limit of weak spatial fluctuation, the effect of correlations is known to produce a correction scaling as the inverse of the coordination number to the Gutzwiller mean-field limit, and therefore MF results are expected to be exact \cite{SchmidtPRL09,BiondiNJP17}.
As pointed out in Ref.~\cite{BiellaPRB18}, however, around second-order critical points correlations diverge, and higher-order correlation schemes should be taken into consideration to properly capture criticality.

Finally, we also stress that linked-cluster expansions explicitly deal with infinite lattice size, while our study is a finite-size one. Moreover, in our lattice, the ratio between the number of sites and the dimension of the lattice $N/d$ is of order one for large lattices, while in the usually defined thermodynamic limit, the number of sites diverges with respect to the dimension.

\subsubsection*{Article structure}
\label{sec:structure}
The article is organized as follows: In Sec.~\ref{sec:model} we introduce the spin model, illustrating its connections with cavity QED models and possible experimental implementations. In Sec.~\ref{sec:mf} we derive the mean-field equations, considering both the case of local and collective dissipation, and the well-studied case of local dissipation only, on which we focus for the main part of the subsequent analysis. 
In Sec.~\ref{sec:liouv} we provide a description of the Liouvillian superoperator and its spectral properties, discussing them in a general case. In particular, in \ref{sec:symmetry} we provide a brief overview of the meaning of Liouvillian symmetries in Lindblad dynamics. In \ref{sec:pt} we introduce
the concept of Liouvillian antigap for $\mathbb{PT}$-symmetric Liouvillians. In \ref{sec:gap} we then compute the closing of the Liouvillian gap and its critical slowing down for the \emph{XYZ} model with local dissipation. 
In Sec.~\ref{Sec:permutational} we introduce the technique that, exploiting permutational symmetry, allows us to calculate various quantities from the steady-state density matrix. 
In Sec.~\ref{sec:validity}, we then compare the mean-field predictions, obtained from analytical solutions, to a numerical study of the quantum model, in the two qualitatively different regimes of the phase diagram. In particular, in Sec.~\ref{sec:critical} we study the properties of the phases across the critical region (paramagnetic phase, critical point, and ferromagnetic phase), while in Sec.~\ref{sec:bcvalidty} we focus on pinpointing the phase transition, in the presence of local dissipation.
In Sec.~\ref{sec:loccol}, we consider the steady-state properties and phase transition in the presence of both local and collective dissipation. Finally, in Sec.~\ref{sec:conclusions} we provide our concluding remarks.

\subsubsection*{Original results}
\label{sec:originalresults}
Before moving to the next sections, we provide a brief summary of the original results obtained in this article:
\begin{itemize}
{\item We derive the phase diagram of the all-to-all dissipative \emph{XYZ} model from both mean-field and quantum steady-state solutions. We find the absence of an antiferromagnetic phase and only one phase transition, from paramagnetic to ferromagnetic phase. A highly-entropic ferromagnetic regime, for high-anisotropy, is found to be qualitatively different from the normal ferromagnet.}
{\item In the presence of both local and collective dissipation, the phase transition is a second-order one, just like the case of local dissipation only (differently from the case of collective dissipation only \cite{LeePRA14}).}
{\item The full quantum Lindblad dynamics converges to the Gutzwiller mean-field steady-state predictions, but, in the anisotropic regime, the discrepancy is much larger than in other regimes, even for $N\simeq 100$ spins; our extensive investigations of various thermodynamic properties are made possible by the use of permutational symmetry in Liouvillian space \cite{ShammahPRA98} and are relevant for state-of-the-art noisy quantum simulators.}

{\item Additionally, we exploit the $\mathbb{PT}$-symmetry of Liouvillians \cite{ProsenPRL12} emerging in spin models whose Hamiltonian has an all-to-all interaction, and the Lindbladian part introduces homogeneous local dissipation processes. We introduce an efficient method to calculate the Liouvillian gap, whose closing marks a dissipative phase transition in the thermodynamic limit, from its symmetric \emph{antigap}, which, as we detail, can be numerically computed much more easily and is a technique that may be applied to other models. }
\end{itemize}

\section{The model and its phase transition}
\label{sec:model}
The Heisenberg model describes the physics of a $d$-dimensional lattice of spins or two-level systems, characterized by nearest-neighbors interaction. 
Its Hamiltonian reads ($\hbar = 1$)
\begin{equation}\label{Eq:Hamiltonian_generic}
\hat{H}=\frac{1}{Z}\sum_{\langle i , j \rangle} \left(J_x \hat{\sigma}^x_i \hat{\sigma}^x_j + J_y \hat{\sigma}^y_i \hat{\sigma}^y_i + J_z \hat{\sigma}^z_i \hat{\sigma}^z_j \right),
\end{equation}
where $Z$ indicates the coordination number, $\langle i , j \rangle$ indicates the sum over nearest-neighbor links, $J_\alpha$ ($\alpha=x,y,z$) represent the coupling strengths of spin-spin interactions, $\hat{\sigma}^\alpha_i$ are the Pauli matrices of the $i$-th spin.
Since we consider $J_x\neq J_y \neq J_z$, we will refer to this anisotropic Heisenberg model as an \emph{XYZ} model.
If such a system weakly interacts with a Markovian environment, its dynamics is captured via a Lindblad master equation \cite{BreuerBookOpen,Haroche_BOOK_Quantum}.
The dissipative part of the dynamics tends to align the spins along the $z$ direction with two different mechanisms.
The first one flips a single spin towards the negative direction of the $z$-axis, with $\gamma$ quantifying the rate of spin-flip processes. 
The second one characterizes the collective loss of one excitation at a rate $\Gamma$. 
The state of the system is thus captured by a density matrix $\rhot$ evolving via
\begin{equation}\label{Eq:Lindblad}
\begin{split}
    \frac{\partial \rhot}{\partial t} & = \mathcal{L} \rhot= - i \left[\hat{H}, \rhot\right] +  \gamma \sum_{j=1}^N \mathcal{D}[\hat{\sigma}^-_j] \rhot \\  & \qquad + \frac{\Gamma}{N -1}  \mathcal{D}[\sum_{j=1}^N \hat{\sigma}^-_j] \rhot,
\end{split}
\end{equation}
where $N$ is the number of two-level systems, $\hat{\sigma}^\pm_j = (\hat{\sigma}^x_j \pm i \hat{\sigma}^y_j)/2$ are the raising and lowering operators for the $j$-th spin, $\mathcal{D}[\hat{A}]$ represents a Lindblad dissipator of the form
\begin{equation}
        \mathcal{D}[\hat{A}]  \rhot  =  \hat{A}\rhot \hat{A}^\dagger - \frac{1}{2} \left(\hat{A}^\dagger  \hat{A} \rhot + \rhot \,  \hat{A}^\dagger  \hat{A} \right), 
\end{equation}
acting on the $j$-th site, and $\mathcal{L}$ is the Liouvillian superoperator. These processes are sketched in Fig.~\ref{fig:xyz}.

If we consider an all-to-all connected model with uniform couplings, i.e., all the spins interact with each other with the same strength, the Hamiltonian in Eq.~\eqref{Eq:Hamiltonian_generic} can be recast as
\begin{equation}\label{Eq:Hamiltonian}
\hat{H}=\frac{1}{2\left(N-1\right)} \left[J_x \left(\hat{S}^{x}\right)^2+J_y\left(\hat{S}^{y}\right)^2+J_z\left(\hat{S}^{z}\right)^2 \right],
\end{equation}
where we have introduced the collective operators $\hat{S^{\alpha}}=\sum_{i=1}^N \hat{\sigma}_i^{\alpha}$ for $\alpha=x, \, y, \,z$. Notice the factor $2$ is due to the fact that in Eq.~\eqref{Eq:Hamiltonian_generic} the sum is over the links while to obtain Eq.~\eqref{Eq:Hamiltonian} we have to sum over the sites.
Moreover, in the all-to-all connected model, the coordination number $Z=N-1$.
The collective dissipation becomes $\mathcal{D}[\sum_j \hat{\sigma}^-_j]=\mathcal{D}[\hat{S}^{-}]$, while the local dissipation cannot be recast in terms of a single collective operator.
In this regard, in the all-to-all connected model, the Hamiltonian and collective dissipation processes will tend to create correlated states, while local dissipation will disentangle them.

However, it is commonly accepted that in a high-dimensional model $d\gg 1$, in the thermodynamic limit fluctuation are suppressed and the correct result should be captured by a mean-field decoupling procedure \cite{MorrisonPRL08,MorrisonPRA08}.
The resulting steady-state density matrix is a tensor product of identical local density matrices.

In this work, we will investigate the phase transition from a paramagnetic phase with no magnetization in the $xy$-plane ($\braket{\hat{\sigma}^x} = \Tr{\hat{\rho}_{\rm ss} \hat{\sigma}^x_j} = 0$ , $\braket{\hat{\sigma}^y} = \Tr{\hat{\rho}_{\rm ss} \hat{\sigma}^y_j} = 0$) to a ferromagnetic phase with finite magnetization in the $xy$-plane ($\braket{\hat{\sigma}^x} \ne 0$ , $\braket{\hat{\sigma}^y} \ne 0$) which is expected to happen in the thermodynamic limit of the \emph{XYZ} model for anisotropic coupling $J_x\neq J_y$ \cite{LeePRL13,JinPRX16,RotaPRB17,OrusNatComm17,BiellaPRA17,CasteelsPRA18, HuybrechtsPRA19}.

\subsection{Collective dissipation: symmetry and relation with superradiant light-matter models}
\label{sec:collective}
Before moving forward to the general case, let us briefly consider the properties of the system in the presence of collective dissipation only, $\Gamma\neq 0$ and $\gamma= 0$ in Eq.~(\ref{Eq:Lindblad}) \cite{MorrisonPRL08,MorrisonPRA08,LeePRA14}.
We then have that the total spin length, 
\begin{eqnarray}
\hat{S}^2&=&\left(\hat{S}^{x}\right)^2+\left(\hat{S}^{y}\right)^2+\left(\hat{S}^{z}\right)^2,
\label{spinlength}
\end{eqnarray}
 is a conserved quantity,
\begin{eqnarray}
\left[\hat{S}^2, \hat{H}\right]&=&\left[\hat{S}^2, \hat{S}^{-}\right]=0,
\label{Eq:S2symm}
\end{eqnarray}
and therefore the presence of conserved quantities implies the existence of several steady states for the Lindbladian dynamics \cite{AlbertPRA14}.
In more physical terms, this indicates that there exist different multiplets, which are eigenstates of $\hat{S}^2$, that are not connected by the dissipative dynamics.
These multiplets are known as Dicke ladders \cite{Dicke54}.

This terminology is inherited from the study of the Dicke model. 
The similarities between the all-to-all connected \emph{XYZ} and Dicke models are both due to mathematical similarities, which will be apparent when exploiting the permutational symmetry, and because this is another benchmark model thoroughly used to investigate both quantum phase transitions and dissipative phase transitions, this time in the field of cavity QED and quantum optics \cite{LeePRA14}.

Describing the collective interaction between an ensemble of two-level systems with a unique photonic field, the Dicke model is known to display superradiant photon emission in the presence of collective dissipation \cite{Bonifacio70,Bonifacio71,Bonifacio75}. Here superradiance refers to the fact that the light emission intensity scales as $N^2$ and occurs on a timescale that shrinks with the size of the system, a macroscopic manifestation of cooperative behavior. Note that this phenomenon does not require any strong coupling between light and matter to occur, so that one can map the light-matter model to an effective spin model that fulfils Eq.~(\ref{Eq:S2symm}), with $\hat{H}=\omega_z \hat{S}^z$, where $\omega_z$ is the resonance frequency.\\

Note that, in the presence of collective coupling only, a Holstein-Primakoff transformation can be performed to map the system to a bosonic model \cite{Lambert04}, whose first-order approximation is valid in the low-excitation regime and is good in the thermodynamic limit.  
The main assumption of coupling only to a collective field is based on the assumption of identical two-level systems (spins) and their identical coupling to the photonic field. When these assumptions are relaxed, intermediate superradiant regimes can still be obtained \cite{Lehmberg70,KesslerPRA12,Buchhold13,Lambert16,DallaTorre16,Kirton17}, resulting from the population of different Dicke ladders \cite{Gegg16,Shammah17}, experimentally verified in solid-state systems \cite{Noe12,Bradac16,Angerer18}. In that case, a bosonic approximation in terms of polaritonic populations can be performed, but only in the low-excitation regime \cite{Shammah17,Cirio19}. In the presence of local incoherent pumping and collective dissipation, the superradiant phase \cite{Dimer07} and steady-state superradiant emission \cite{Meiser10a} have been proposed and observed in cavity QED setups with atomic clouds \cite{Baumann10,Bohnet12}. Similarly, trapped ions and atomic lattices provide the opportunity to engineer long-range interactions and dissipation \cite{Niederle16,Gelhausen17}, relevant also for the implementation of the anisotropic Heisenberg models \cite{Bermudez17b}.

\subsection{Experimental implementations}
\label{sec:experiments}
We envision that the predictions that will be detailed hereafter can be observed in experiments with noisy quantum simulators and long-range interaction, based on a broad variety of platforms: atomic clouds \cite{ViteauPRL12}, Rydberg atoms \cite{NguyenPRX18,LeePRA11,QianPRA12,ChanPRA2015,GlaetzlePRL15}, trapped ions \cite{Russomanno17,Zhang2017,Davoudi19,RamosEPJD19}, as well as in solid state \cite{Noe12,Iemini17}, e.g., in superconducting circuits \cite{YouNat11,HouckNatPhys12,TsomokosNJP08,Lambert09,Kakuyanagi16,MarkovicPRL18} and especially in hybrid superconducting systems \cite{Angerer18}, where a bosonic field mediates the effective spin-spin interactions. 
Indeed Ref.~\cite{NguyenPRX18} shows the feasibility of investigating exactly the all-to-all connected \emph{XYZ} model in Rydberg atoms. Probing the dissipative regime here studied only requires implementing a weak-coupling interaction with an additional cavity mode allowing for dispersive measurement of the radiated field.
Trapped ions provide another platform on which to engineer long-range spin interactions \cite{Zhang2017,Davoudi19, RamosEPJD19} and already allow one to investigate dissipative phase transitions with tens of two-level systems, which can also be locally manipulated \cite{Garttner17}.

Superconducting circuit elements and condensed matter magnetic degrees of freedom can be plugged together to implement hybrid quantum systems. One such example is provided by a collection of nitrogen vacancies (NV) or color centers in diamond interacting with the magnetic field controlled by a superconducting resonator. This platform offers the advantage of large $N$ spins, actually implementing a good approximation of the thermodynamic limit since $N \approx 10^{12}$--$10^{16}$ there, and physical conditions that allow to explore various regimes of both collective and local dissipation. The former is determined by the superconducting resonator quality factor, the latter by the intrinsic impurities of the condensed matter system and couplings to the crystal lattice. In these systems, superradiant light emission has been recently observed \cite{Bradac16,Angerer18}, as well as steady-state bistability and critical slowing down \cite{Angerer17}. In the bad-cavity regime, the cavity mode decay allows an adiabatic elimination of the bosonic degree of freedom, allowing the implementation of effective spin Hamiltonians, while tuning spin sub-ensembles in and out of resonance allows to vary $N$ and thus study system-size scaling \cite{Angerer18}.

\section{Mean field treatment}
\label{sec:mf}
Solving the Lindblad master equation \eqref{Eq:Lindblad} by assuming a Gutzwiller ansatz for the density matrix results in the following set of mean-field (MF) equations,

\begin{widetext}
\begin{subequations}
\label{Eq:Collective}
\begin{eqnarray}
\label{Eq:CollectiveX} \partial_t \braket{\hat{\sigma}^x}&=&
2\left(J_y-J_z\right)\braket{\hat{\sigma}^y} \braket{\hat{\sigma}^z}
-\frac{\tilde\gamma}{2}\braket{\hat{\sigma}^x}
+\frac{\Gamma}{2}\braket{\hat{\sigma}^x} \braket{\hat{\sigma}^z},\\
\label{Eq:CollectiveY}\partial_t \braket{\hat{\sigma}^y} &=&
2\left(J_z-J_x\right)\braket{\hat{\sigma}^x} \braket{\hat{\sigma}^z}
-\frac{\tilde\gamma}{2} \braket{\hat{\sigma}^y}
+\frac{\Gamma}{2}\braket{\hat{\sigma}^y} \braket{\hat{\sigma}^z},\\
\label{Eq:CollectiveZ}\partial_t \braket{\hat{\sigma}^z}&=&
2\left(J_x-J_y\right)\braket{\hat{\sigma}^x} \braket{\hat{\sigma}^y}
-\tilde\gamma(\braket{\hat{\sigma}^z}+1)
-\frac{\Gamma}{2}\left(\braket{\hat{\sigma}^x}^2+\braket{\hat{\sigma}^y}^2\right),
\end{eqnarray}
\end{subequations}
\end{widetext}
having defined $\tilde\gamma=\gamma+\Gamma/(N-1)$ and $\braket{\hat{\sigma}^\alpha}$ the single-site approximation of the Pauli matrix expectation values, with $\alpha=x,\;y,\;z$.

To perform the mean-field analysis we model the all-to-all coupled spin system as a $d$-dimensional system. Every time we add a spin the dimension of the system is also increased by one. This implies that a $d$-dimensional system consists of $d$ spins and that infinite dimensions are reached when the system has an infinite amount of spins. 

Firstly, in this section, we will derive the solutions of the MF equations (\ref{Eq:Collective}) and study their singularities to predict the location in phase space of the phase transition and its characteristics. 
In Sec.~\ref{sec:validity} we will then test if mean-field theory becomes exact in infinite dimensions, i.e. infinite number of spins, by comparing with the exact results for increasing number of spins. This analysis will provide a benchmark for spin models on the correctness of mean-field theory in dissipative systems, beyond results found for dissipative spin-boson models \cite{KirtonNJP2017, Kirton19rev}.

Equation~\eqref{Eq:Collective} can be easily solved numerically. However, from its inspection we can retrieve some insight on the interplay of processes in the dynamics. Its analytical solution, even in the steady state ($\partial_t \braket{\hat{\sigma}^\alpha}=0$), is complicated by the inclusion of collective emission. This process introduces dissipative nonlinear terms that, for Eq.~(\ref{Eq:CollectiveX}) and Eq.~(\ref{Eq:CollectiveY}) are similar to the Hamiltonian ones, hinting at the fact that they contribute to entanglement generation in the dynamics; for Eq.~(\ref{Eq:CollectiveZ}), the symmetry present in the Hamiltonian terms is instead broken by the nonlinear term in $\propto \Gamma \left(\braket{\hat{\sigma}^x}^2+\braket{\hat{\sigma}^y}^2\right)$, which, moreover, cannot be simplified in terms of $\braket{\hat{\sigma}^z}^2$, in the presence of local dissipation, since the spin length, Eq.~(\ref{spinlength}), is not preserved.    

\begin{figure}
    \centering
    \includegraphics[width=\linewidth]{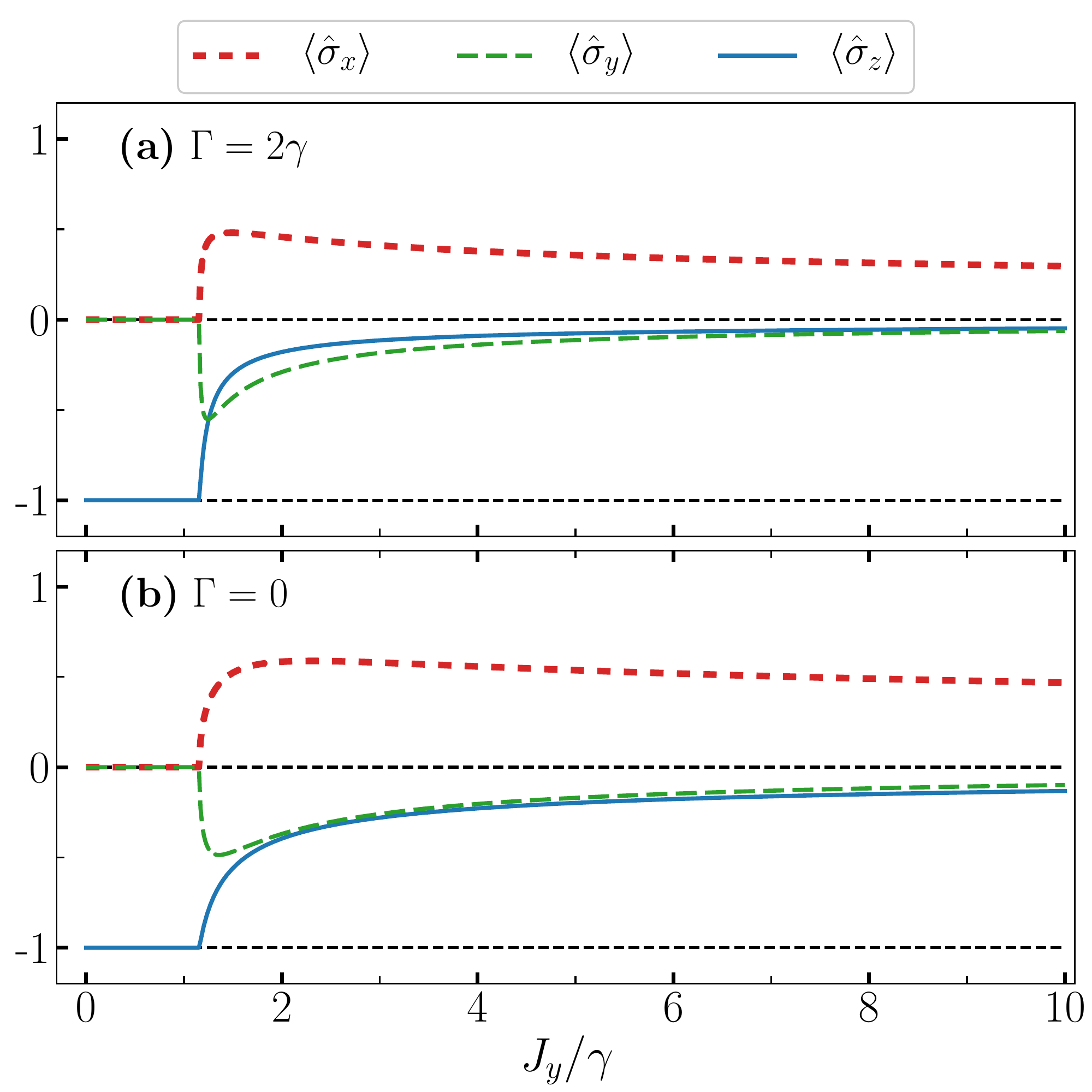}
    \caption{Steady-state solution of the mean-field equations~\eqref{Eq:Collective} in the case $\Gamma=2 \gamma$ [panel (a)] and in the case $\Gamma=0$ [panel (b)], having fixed the value $\left(\gamma + \Gamma\right) = 1$. 
    The parameters used here are $J_x/(\gamma + \Gamma)=0.6$, $J_z/(\gamma + \Gamma)=1$ and $N\to\infty$. The horizontal black dashed lines correspond to $\braket{\hat\sigma_\alpha}=0,-1$.
    }
    \label{fig:MF_collective_vs_local}
\end{figure}

We plot the MF solution to Eq.~\eqref{Eq:Collective} in Fig. \ref{fig:MF_collective_vs_local} in the case $\Gamma=2 \gamma$ [panel (a)] and in the case $\Gamma=0$ [panel (b)].
The total dissipation $\left(\gamma + \Gamma\right)$ is kept fixed.
We notice that both MF solutions predict a second-order phase transition and that the value of $J_y$ triggering the phase transition is the same in both cases.
However, the two plots exhibit a different dependence of the mean values $\braket{\hat{\sigma}^\alpha}$ on $J_y$, with $\alpha=x, \, y, \,z$. In the presence of local and collective dissipation [panel (a)], the transition appears to be sharper than in the presence of local dissipation only [panel (b)].

\subsection{Local dissipation only}
\label{sec:local}
We will now focus our analysis on the case $\Gamma=0$ in Eq.~(\ref{Eq:Collective}), which was extensively investigated in Refs.~\cite{JoshiPRA13,LeePRL13,JinPRX16,RotaPRB17,OrusNatComm17,BiellaPRA17,CasteelsPRA18,RotaNJP18, HuybrechtsPRA19} in lower dimensions and in Ref.~\cite{LeePRA14} in infinite dimension. The MF equations of motion are
\begin{subequations}
\begin{eqnarray}
        \label{mfeqX}\partial_t\langle \hat{\sigma}^{x}\rangle &=& -\frac{\gamma \langle \hat{\sigma}^{x}\rangle}{2} + 2 \left(J_y -J_z\right) \langle \hat{\sigma}^{y}\rangle \langle \hat{\sigma}^{z}\rangle ,\label{mfeqX}\\
        \label{mfeqY}\partial_t\langle \hat{\sigma}^{y}\rangle &=& -\frac{\gamma \langle \hat{\sigma}^{y}\rangle}{2} + 2 \left(J_z -J_x\right) \langle \hat{\sigma}^{x}\rangle\langle \hat{\sigma}^{z}\rangle ,\label{mfeqY}\\
        \label{mfeqZ}\partial_t\langle \hat{\sigma}^{z}\rangle &=& -\gamma\left(\langle \hat{\sigma}^{z}\rangle + 1\right)  + 2 \left(J_x -J_y\right) \langle \hat{\sigma}^{x}\rangle\langle \hat{\sigma}^{y}\rangle .\label{mfeqZ}
\end{eqnarray}
\label{mfeq}
\end{subequations}

We notice that Eqs.~(\ref{mfeqX}-\ref{mfeqZ}), although nonlinear, are analytically solvable for the steady state. They only contain nonlinear \emph{homogeneous} terms, and one can thus obtain $\langle \hat{\sigma}^{z}\rangle_{\rm ss}$ exactly.

\begin{figure}[!h]
  \centering
    \includegraphics[width=0.5\textwidth]{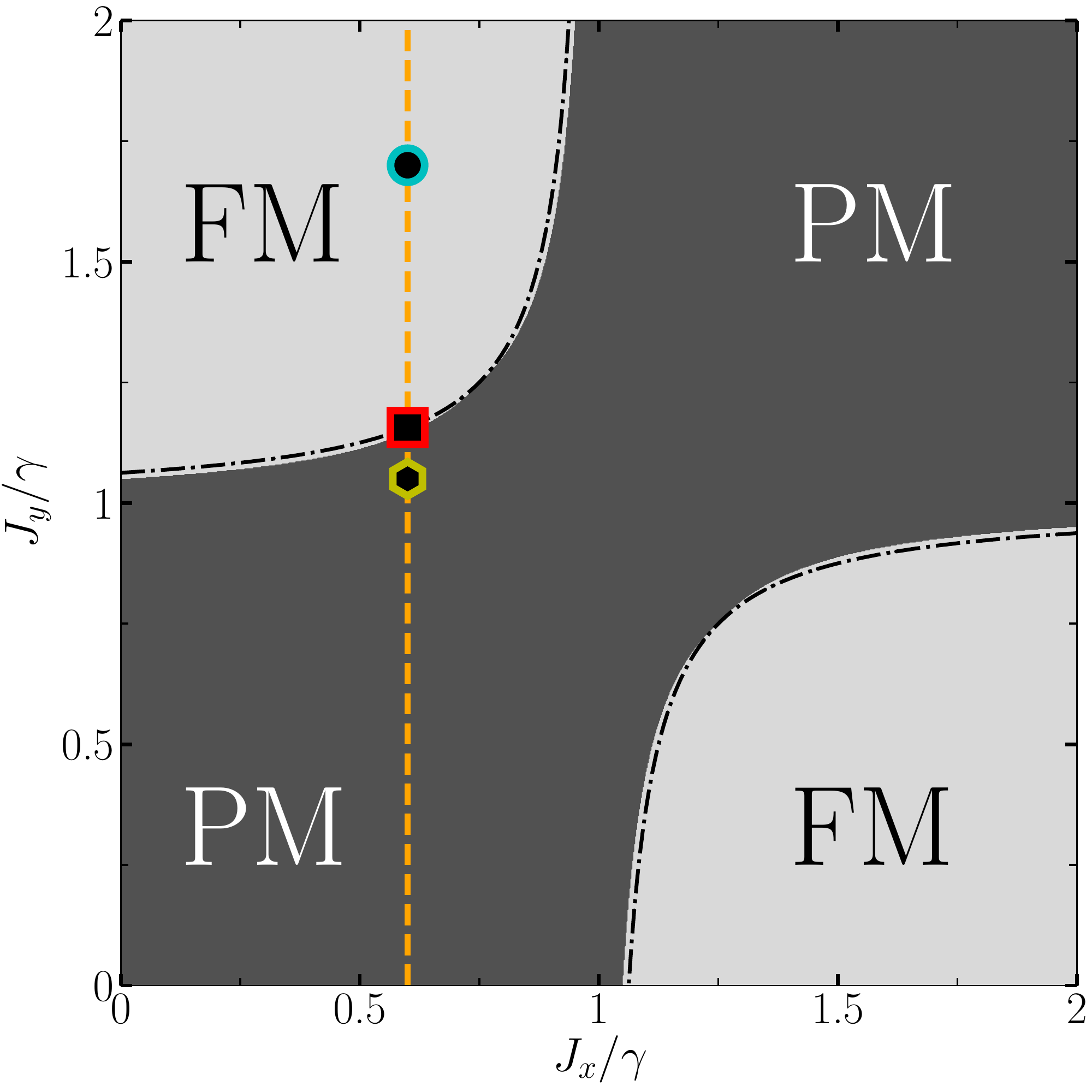}

      \caption{Phase diagram for local dissipation only, where $\Gamma=0$ and $J_z/\gamma = 1$. The phases are determined from the intersection in the bimodality coefficient curves for in the $x$ and $y$ direction for $N =50$ and $N = 60$, i.e. the transition from a paramagnetic phase (PM) to a ferromagnetic phase (FM) in the $xy$-plane. The black dash-dotted curves show where the transition takes place in the mean-field approximation, while the background color defines the PM (dark grey) and FM (light grey) regions from calculations using the bimodality coefficient in the full quantum model. The orange vertical dashed line is located at $J_x = 0.6\gamma$ and shows the cut that will be used in the next figures to characterize the phase transition. The three points on the cut $J_x/\gamma = 0.6$ indicate the values of $J_y/\gamma$ which will be used for bench-marking the MF with the full quantum solutions: $J_y/\gamma=1.1$, in the PM phase (hexagon with yellow contour), at criticality, $J_y/\gamma=1.15625$ (square with red contour), and at $J_y/\gamma=1.7$ in the moderately anisotropic FM region (circle with a cyan contour).} \label{fig:bcpd}
\end{figure}

We study the mean-field phase diagram through an instability analysis analogous to the one performed for the nearest neighbor \emph{XYZ} Hamiltonian \cite{LeePRL13}. We determine the instability of the paramagnetic phase in the $xy$-plane to a $d$-dimensional perturbation with wave vector $\Vec{k}$. Due to the all-to-all connected structure, the perturbations with wave vector $\Vec{k} = \left(k_1, k_2, ..., k_d\right)$ are restricted by $k_l$ only being able to attain the values $0$ and $\pi$. For such analysis the presence of an antiferromagnetic phase is nonphysical for any value of the coupling parameters. Hence, the mean-field phase diagram consists only of a paramagnetic phase and a ferromagnetic one. The latter is present when the condition
\begin{equation}
    -\frac{\gamma^2}{16} > \left(J_x - J_z\right)\left(J_y - J_z \right),
    \label{eqinst}
\end{equation}
is fulfilled. The absence of an antiferromagnetic phase in this all-to-all connected model can be expected. Each spin is connected to every other spin in the system, and no unique spatial structure is present for this type of interaction. It is impossible for the spins to take alternating directions with respect to their neighbors. The results of this instability analysis lead to the phase diagram shown in Fig.~\ref{fig:bcpd}, where the black dash-dotted curves show the transition boundary between both phases according to the mean-field approximation.

We will now proceed to study the dynamics in the full quantum formalism. Unveiling its symmetries, and especially exploiting permutational symmetry numerically, we will be able to calculate several physical properties of the steady-state density matrix. After that, we will be in a position to precisely perform a comparative analysis with respect to the mean-field predictions derived from the solutions obtained here. 

\section{Liouvillian spectrum and phase transitions}
\label{sec:liouv}
Let us briefly revise the properties of Liouvillians, and their relation to phase transitions \cite{MingantiPRA18_Spectral}.
Given any Liouvillian $\LL$, we can introduce its eigenvalues $\lambda_i$ and eigenmatrices $\eig{i}$, defined via the relation
\begin{equation}
\LL \eig{i}=\lambda_i \eig{i}.
\end{equation}
From a numerical point of view, and for the model under consideration, we can obtain the eigenvalues and eigenmatrices of the Liouvillian by diagonalizing the matrix representation of $\mathcal{L}$,
\begin{equation}
\begin{split}
    \mathcal{L} &= -i\left(\hat{H} \otimes \mathbb{1}-\mathbb{1} \otimes \hat{H}^{\mathrm{T}}\right)+ \\ 
     & \quad \gamma \sum_j^N \left( \hat{\sigma}^-_j \otimes \hat{\sigma}^-_j-\frac{\hat{\sigma}^+_j\hat{\sigma}^-_j \otimes \mathbb{1}+ \mathbb{1} \otimes \hat{\sigma}^+_j\hat{\sigma}^-_j}{2} \right) + 
     \\
     &\quad \Gamma \left( \hat{S}^- \otimes \hat{S}^- -\frac{\hat{S}^+\hat{S}^- \otimes \mathbb{1}+ \mathbb{1} \otimes \hat{S}^+\hat{S}^-}{2}\right)
     . 
\end{split}
\end{equation}
Here $\hat{A}^T$ represents the transpose of the operator $\hat{A}$.
Since this Liouvillian is not Hermitian, in general its eigenvalues $\lambda_i$ need not to be real.
It can be proved that for any Liouvillian, given an eigenvalue $\lambda_i$ whose eigenmatrix is $\eig{i}$, there exist a $\eig{i}^\dagger$ whose eigenvalue is $\lambda_i^*$ \cite{MingantiPRA18_Spectral}.
Therefore, the eigenvalues are symmetrically distributed with respect to the real axis, as shown in Fig.~\ref{fig:spectra}.
Moreover, they are characterized by $\Re{\lambda_i}\leq 0$.
We order the eigenvalues $\lambda_i$ in such a way that $|\Re{\lambda_0}|<|\Re{\lambda_1}| <\dots <|\Re{\lambda_n}| < \dots$, i.e., the eigenvalues are ordered by their real part.
In this regard, the steady state, that is the density matrix $\sss$ such that $\LL \sss=0$, is the eigenmatrix of the Liouvillian associated to the zero eigenvalue.
The real part of the eigenvalues describes the relaxation towards the steady-state of a generic matrix, while the complex part describes the oscillatory processes which may take place.
A fundamental role is played by $\eig{1}$, that is the eigenmatrix associated to the smallest eigenvalue $\lambda_1$, which describes the slowest relaxation scale towards the steady-state.
A phase transition takes place in the thermodynamic limit when $\lambda_1$ becomes exactly zero, both in its real and imaginary parts.
For any finite size of the system under consideration, however, $\lambda_1\neq 0$.
Nevertheless, the study of $\lambda_1$ and $\eig{1}$ provides much useful information about the scaling and nature of the transition \cite{VicentiniPRA18}.

\subsection{Symmetry breaking and phase transition}
\label{sec:symmetry}
The Lindblad master equation~\eqref{Eq:Lindblad} is invariant under a $\pi$-rotation of all the spins around the $z$-axis ($\hat{\sigma}_i^{x} \to - \hat{\sigma}_i^{x}$, $\hat{\sigma}_i^{y} \to - \hat{\sigma}_i^{y}$ $\forall i$). 
Thus, the system admits a $\mathcal{Z}_2$ symmetry, that is, there is a superoperator $\mathcal{Z}_2$ such that
\begin{equation}\label{Eq:symmetry}
\mathcal{Z}_2 \rhot = \prod_{j=1}^N\exp{\left(-i \pi \hat{\sigma}^z_j\right)}\rhot \prod_{j'=1}^N\exp{\left(+i \pi \hat{\sigma}^z_{j'}\right)},
\end{equation}
and one can verify that $[\LL, \mathcal{Z}_2]=0$.
While in a Hamiltonian system the presence of a symmetry implies a conserved quantity, this is not always the case for Liouvillian symmetries \cite{AlbertPRA14,BaumgartnerJPA08}.
A symmetry of an out-of-equilibrium system, however, implies that the steady-state cannot have an arbitrary structure. 
In our case, $\sss$ must be an eigenmatrix of $\mathcal{Z}_2$, such that $\mathcal{Z}_2 \sss \propto \sss$.
In turn, this means that, for any finite size system $\braket{\hat{\sigma}^x_i}=\braket{\hat{\sigma}^y_i}=0$ for all sites $i$.

The symmetry breaking takes place when, in the thermodynamic limit, $\lambda_1=0$ allows to have two steady states with nonzero and opposite magnetization.
We thus expect to observe a second-order phase transition associated to this symmetry breaking of $\mathcal{Z}_2$ \cite{MingantiPRA18_Spectral}.
For a finite-size system, $\lambda_1 \neq 0$, such symmetry breaking cannot be directly witnessed. 
However, the precursors of the phase transition can be inferred both via spectral analysis and via an extensive study of the scaling of observables (see the discussion in Sec.~\ref{Sec:permutational}).

\subsection{$\mathbb{PT}$-symmetry and Liouvillian antigap}
\label{sec:pt}
There exists a class of non-Hermitian Hamiltonian systems which are invariant under the composition of unitary (parity $\mathcal{P}$) and anti-unitary (time-reversal $\mathcal{T}$) transformations: the $\mathcal{PT}$-symmetry \cite{El-GanainyNature2018,MirieaarScience19,Ozdemir2019}.
This $\mathcal{PT}$-symmetry cannot be directly extended to the Liouvillian case, due to the dissipative nature of the contractive dynamics \cite{Scheel2018}.
However, certain systems admit a $\mathcal{PT}$-symmetric transformation once a shift parallel to an average damping rate is added to $\LL$ \cite{ProsenPRL12}.
Therefore, the $\mathbb{PT}$-symmetry of $\LL$ is not a superoperator symmetry in the sense of Eq.~\eqref{Eq:symmetry} (that is, it does not describe a property of the steady state).
Instead, it is a spectral property related to the emergence of a reflection symmetry of the eigenvalues in the complex plane, i.e. introducing a dihedral ($D_2$) symmetry. 
Indeed, there exist a real number $\eta>0$ such that, for all the eigenvalues $\lambda_i$, there exist a $\lambda_j =  -2 \eta + \lambda_i$.
This can be easily visualized by plotting the eigenvalues of the Liouvillian in the complex plane $\lambda_j=x_j+ i \, y_j$.

\begin{figure*}[!ht]
  \centering
    \includegraphics[width=0.8\textwidth]{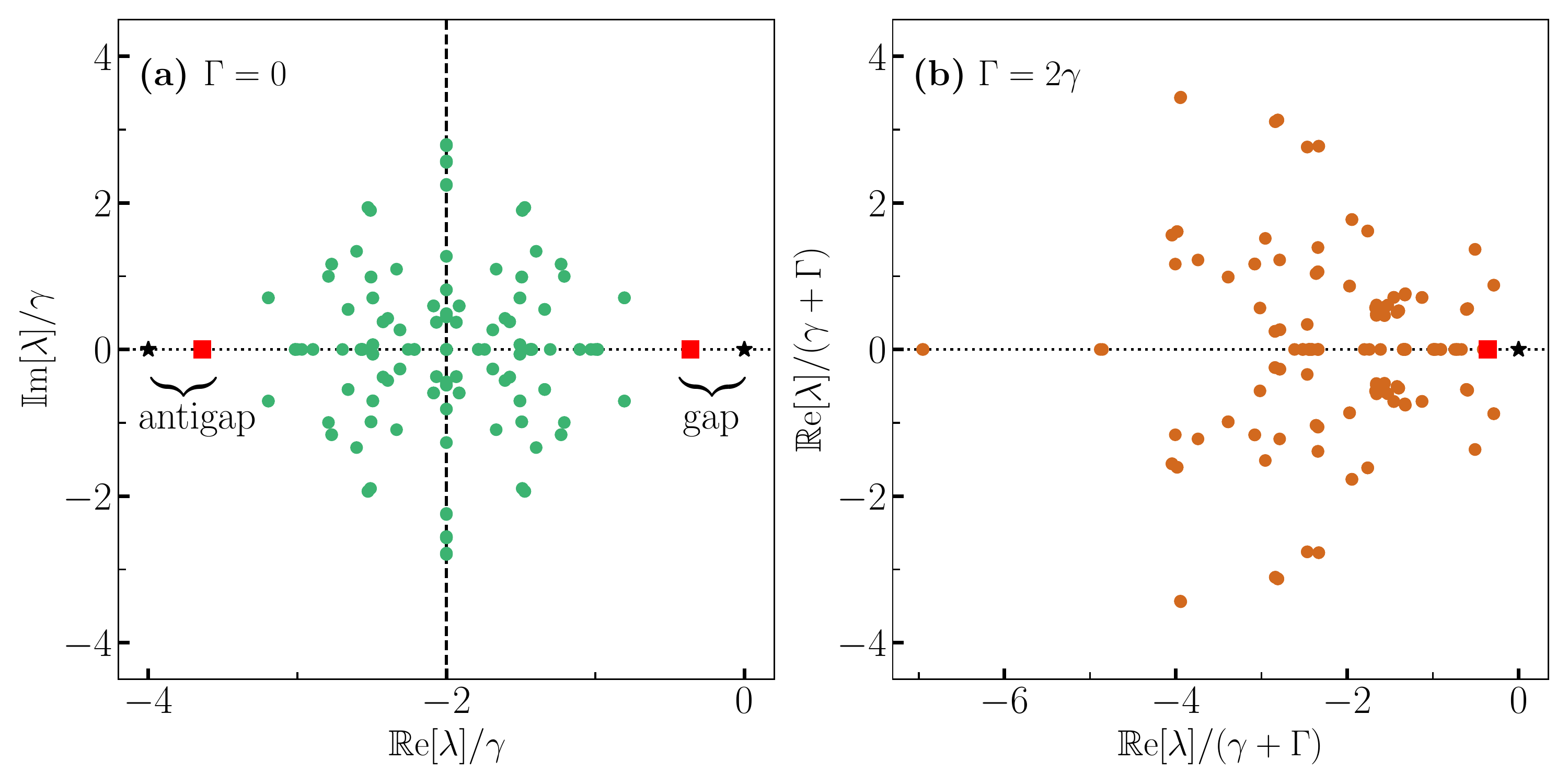}
      \caption{
      Liouvillian spectrum for the dissipative \emph{XYZ} model with local dissipation [panel (a)], with the system parameters as specified in Fig. \ref{fig:MF_collective_vs_local} panel (b), and both local and collective dissipation [panel (b)], with the system parameters as specified in Fig. \ref{fig:MF_collective_vs_local} panel (a). Here $N=4$ and we choose $J_y=J_z$. We  mark $\lambda_0$ and $\lambda_M$ with a black star and a red square, respectively. All other eigenvalues $\lambda_i$ are marked by circles. (a) The $\mathbb{PT}$-symmetry of the Liouvillian with only local dissipation is visible by the additional plane symmetry (vertical dashed line) of the eigenvalues (green circles). The Liouvillian gap and the Liouvillian \emph{antigap} of the $\mathbb{PT}$-symmetric model are highlighted. (b) The Liouvillian spectrum with local and collective dissipation, showing no $\mathbb{PT}$-symmtery. }
\label{fig:spectra}
\end{figure*}

The $\mathbb{PT}$-symmetry results in a reflection symmetry of the eigenvalues with respect to a line $x=-\eta$ parallel to the imaginary axis \cite{ProsenPRL12,ProsenPRA12,vanCaspel18}. The spectrum of the dissipative all-to-all connected \emph{XYZ} spin model is shown in Fig.~\ref{fig:spectra}, setting $N=4$, $J_x=0.6J_z$ and $J_y=J_z$. In Fig.~\ref{fig:spectra}(a) we consider the case of homogeneous local dissipation, $\Gamma=0$ in Eq.~(\ref{Eq:Lindblad}),  and for comparison, the case of collective and local dissipation is shown in Fig.~\ref{fig:spectra}(b), $\Gamma=2\gamma$ in Eq.~(\ref{Eq:Lindblad}), showing instead no additional symmetry in the spectrum. We have verified that the absence of $\mathbb{PT}$-symmetry occurs also in the case of collective dissipation only, $\gamma=0$, $\Gamma\neq0$. Similarly, also in the case of local dephasing and local pumping, the Liouvillian spectrum of the model displays the additional dihedral symmetry typical of $\mathbb{PT}$-symmetry.

To clarify the discussion, let us consider a $\mathbb{PT}$-symmetric Liouvillian  with $(M+1)$ eigenvalues.
Therefore, there exists an eigenmatrix $\eig{M}$ whose eigenvalue is $\lambda_M$, which is the symmetric counterpart of $\sss$.
Since $\lambda_0=0$ and $\lambda_M=-2 \eta$, we can directly access the value of $\eta$. 
Similarly, we can define the eigenmatrix $\eig{M-1}$ which mirrors $\eig{1}$, and an ``antigap'' $\lambda_{M-1}$, such that $\lambda_{M-1} - \lambda_{M} = \lambda_1$.
This property allows for an easier numerical computation of the gap and associated $\eig{1}$.
Indeed, if one is interested in computing only a few eigenvalues of the Liouvillian, one could resort to an iterative diagonalization method, based on Krylov subspaces.
This method works extremely well for large-magnitude eigenvalues.
However, if one is interested in computation of small eigenvalues, this method performs worse.
Indeed, one has to invert the matrix $\LL$, so that the eigenvalues of smallest magnitude become the most relevant ones.
Moreover, for non-Hermitian matrices, this method is known to be unstable \cite{Nation15}.
Knowing that the Liouvillian is $\mathbb{PT}$-symmetric (and knowing $\eta$) can mitigate these numerical problems: by considering the shifted Liouvillian $\LL'=\LL + 2 \eta \, \mathbb{I}$, the steady state is characterized by $\lambda_0'=2\eta$ and $\lambda_1'=2\eta - \lambda_1$, where $\mathbb{I}$ is the identity matrix.

In a \emph{XYZ} spin system, a sufficient condition to have this $\mathbb{PT}$-symmetric behavior is to have dissipation only on the border of the chain \cite{ProsenPRA12}.
This condition is trivially satisfied for the all-to-all connected \emph{XYZ} spin model, since all spins are at the border of the system.

\begin{figure}[!ht]
  \centering
    \includegraphics[width=0.5\textwidth]{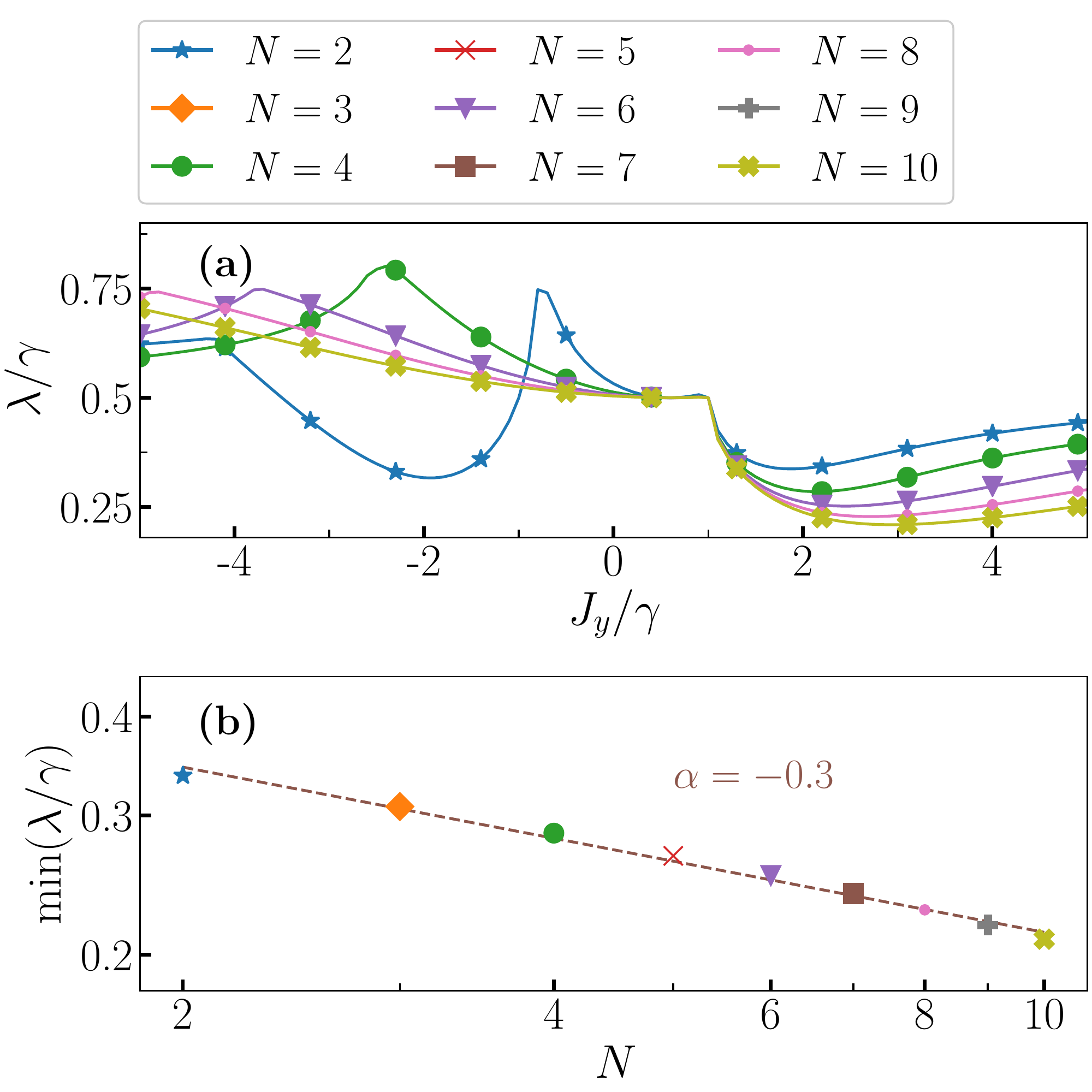}
      \caption{
      Study of the Liouvillian gap, in units of the local dissipation rate, $\gamma$, and its critical slowing down for the dissipative \emph{XYZ} model with local dissipation only. The system parameters are chosen as specified in Fig. \ref{fig:MF_collective_vs_local} panel (b). (a)
      The Liouvillian gap, $\lambda$, is plotted as a function of $J_y/\gamma$ for various system sizes, $N=2,\dots,10$. The markers are only a guide for the eye (101 points have been calculated for value of $N$). (b) The minimum of the Liouvillian gap, normalized by $\gamma$, for each of the curves in the top panel is plotted as a function of the system size $N$ in a log-log plot, showing a linear scaling of the Liouvillian gap typical of phase transition ($\lambda \propto N^\alpha$, with exponent $\alpha=-0.3$) leading to a critical slowing down in the thermodynamic limit.}
\label{fig:gaps}
\end{figure}

\subsection{Closing of the Liouvillian gap: critical slowing down}
\label{sec:gap}
The occurrence of the phase transition in dissipative quantum systems evolving under a Lindblad dynamics is marked by the closing of the Liouvillian gap. A study of the closing of the Liouvillian gap is reported in Fig.~\ref{fig:gaps}. Exploiting the $\mathbb{PT}$-symmetric \emph{antigap} method introduced in Sec.~\ref{sec:pt}, we can even extract the gap of the Liouvillian matrix from the full $4^N\times4^N$ Liouvillian space representation, for small system sizes.    

In Fig.~\ref{fig:gaps}(a), the Liouvillian gap, $\lambda$, is calculated as a function of $J_y$ (normalizing both quantities by a fixed value of $\gamma$), for various system sizes, $N$, also setting $J_z=\gamma$, $J_x=0.6\gamma$. In panel (a), it is visible how the gap tends to close abruptly after $J_y/\gamma\simeq1$, with a minimum that shifts toward $J_y/\gamma=3$ for $N=10$. No critical behavior is observed for small or negative values of $J_y/\gamma$, hinting at the absence of an antiferromagnetic phase. Beyond the FM to PM phase transition predicted by the mean field, and here corroborated by the abrupt decrease of $\lambda$, for larger values of $J_y/\gamma$, we see that the Liouvillian gap again increases. However, for larger values of $N$, the magnitude of this effect is diminished. This aspect already provides hints to the fact that a second-order dissipative phase transition is occurring, as these are the only ones characterized by a closing of the gap over an extended region of the control parameter \cite{MingantiPRA18_Spectral}.  

A study of the critical slowing down is performed in Fig.~\ref{fig:gaps}(b), where the minimum of the Liouvillian gap for each curve of panel (a) is plotted against the system size $N$ in a log-log plot, showing an excellent fit by a power law $\text{min}(\lambda/\gamma)=\beta N^\alpha$ with exponent $\alpha=-0.3$.

Having demonstrated via spectral analysis the presence of the paramagnetic-to-ferromagnetic phase transition and the absence of an antiferromagnetic regime, we consider now the properties of the steady-state density matrix.

\section{Exploiting the permutational symmetry: Calculation of physical quantities}
\label{Sec:permutational}
From a computational point of view, the numerical solution of the master equation \eqref{Eq:Lindblad} is a formidable task when considering extended lattices.
 The density matrix for $N$ spins lives in a $2^N$-dimensional Hilbert space. If one were interested only in the Hamiltonian unitary dynamics, the Hilbert space dimension reduces to $(N+1)$, at most, using the basis of collective spin states. These are the Dicke states $\ket{j,m}$, where $j$ is the cooperation number of the collective spin length and $m$ its projection along one of the axes ($0\leq j\leq \frac{N}{2}$ and $|m|\leq j$, both are integer or semi-integer numbers).

 However, in general, considering local dissipation to separate environments in the Lindblad master equation~(\ref{Eq:Lindblad}), requires storing a matrix of size $4^N\times 4^N$. If one assumes that each spin dissipates at the same rate $\gamma$, the system possesses permutational symmetry also in Liouvillian space \cite{ShammahPRA98}. The presence of local dissipation connects spin multiplets with different cooperation number $j$. The description of the dynamics can still be performed using only $O(N^3)$ computational resources, as detailed in Ref.~\cite{ShammahPRA98}. 

For numerical purposes, one of the key features of the density matrix of the collective system is its block-diagonal structure, arising from the fact that permutational invariance forbids coherences between matrix elements ${\rho}_{j,m;j'm'}=\langle j',m'| \hat{\rho}|j,m\rangle$ for $j\neq j'$. This allows to consider the matrix $\hat{\rho}_{j,m,m'}=\bigoplus_{j=j_\text{min}}^{N/2}\hat{\rho}_{j}$, where each block $\hat{\rho}_{j}$ has dimension $(2j+1)\times(2j+1)$ through which $m$ and $m'$ run, and $j_\text{min}$ is either 0 or $1/2$ for even or odd number of spins, respectively. There are thus $O(N^2)$ matrix elements in each block for $O(N)$ blocks, making the number of elements required to characterize $\hat{\rho}$ only $O(N^3)$. 
This matrix representation exploits the fact that, for each block $\hat{\rho}_j$, there are actually $d_j^{(N)}$ identical blocks with the same matrix elements \cite{ShammahPRA98,Novo13}, 
\begin{eqnarray} d_j^{(N)}&=&(2j+1)\frac{N!}{\left(\frac{N}{2}+j+1\right)!\left(\frac{N}{2}-j\right)!}.
\label{eq:degeneracy}
\end{eqnarray}
 When one calculates collective properties based on operators expectation values, $\langle A \rangle = \text{Tr}[A\hat{\rho}]$, the average over identical blocks is implicit due to the linearity of the trace: one can neglect the degeneracy, Eq.~(\ref{eq:degeneracy}), and directly compute the expectation values. 

However, in order to calculate quantities obtained from the trace of {\emph{nonlinear}} functions of the density matrix, $f[\hat{\rho}]$, such as the Von Neumann entropy, $S[\hat{\rho}]=\text{Tr}[\hat{\rho}\text{log}(\hat{\rho})]$, or the purity, $\mu[\hat{\rho}]=  \text{Tr}[\hat{\rho}^2]$, it is necessary to account for the degeneracy of each block of such block-diagonal density matrix, weighting the contribution of each degenerate block with the factor $ d_j^{(N)}$ of Eq.~(\ref{eq:degeneracy}), 
\begin{eqnarray}
f[\hat{\rho}_{j,m,m'}]&=&\sum_{j=j_\text{min}}^{N/2}d_j^{(N)} \text{Tr}[f[\hat{\rho}_{j}/d_j^{(N)}]].
\end{eqnarray}

\subsection{Spin structure factor and $z$-magnetization}
\label{sec:spinstructure}
To identify the possible agreement of the mean-field theory with the exact numerical solutions we will study the order parameter of the system. Due to the $\mathcal{Z}_2$-symmetry present in the system we cannot rely on the magnetization in the $x$- and $y$-direction. As a result we study the steady-state spin structure factor, which is calculated as follows
\begin{equation}
    S^{\alpha\beta}\left(\textbf{k}\right) = \frac{1}{N(N-1)}\sum_{j\neq l}e^{i\textbf{k}\cdot(\textbf{j}-\textbf{l})}\langle\hat{\sigma}_{j}^{\alpha}\hat{\sigma}_{l}^{\beta}\rangle,
\label{Eq:sxx}
\end{equation}
where $\alpha$, $\beta = x$ or $y$ and where $\langle\hat{\sigma}_{j}^{\alpha}\hat{\sigma}_{l}^{\beta}\rangle=\text{Tr}[\hat{\sigma}_{j}^{\alpha}\hat{\sigma}_{l}^{\beta}\hat{\rho}_{\rm ss}]$. It contains information on the orientation of the spins with respect to each other. Ferromagnetic order is present in the $xy$-plane if the steady-state spin structure factor in the $x$-direction or (and) the $y$-direction is different from zero.

We note that in Eq.~(\ref{Eq:sxx}) the spin structure factor is defined without the contribution of the self-energies, i.e. the sum over the sites considers only different spins. We can thus calculate these quantities even for permutational-symmetric systems, subtracting the single-site contributions to the total second moments.

If we consider $S^{xx}\left(\textbf{k}=0\right)$ or $S^{yy}\left(\textbf{k}=0\right)$ (and from now on we will drop the $\textbf{k}=0$), they predict a ferromagnetic phase when they are nonzero and a paramagnetic phase when they are both equal to zero. Besides being able to identify the phase we are also interested in the quantitative agreement of the mean-field theory with the exact solutions. To this end, we will also study the $z$-magnetization in the steady state, $M_z=\text{Tr}[\hat{\rho}_{\rm ss}\hat{S}^z]/N$, which can be readily calculated without the limitations of the $\mathcal{Z}_2$-symmetry. 

\subsection{Von Neumann entropy}
\label{sec:vn}
The study of the Von Neumann entropy of the steady state is an interesting extension of our previous analysis, since in standard thermodynamics a second-order phase transition is associated to a change in the entropy of the system. 
The Von Neumann entropy reads 
\begin{equation}
    S = -\sum_i p_i \text{log}\left(p_i\right),
    \label{vn}
\end{equation}
with $p_i$ the eigenvalues of the density matrix. It can thus provide information on the mixed nature of the steady-state density matrix, $\hat{\rho}_{\rm ss}$. Usually in many-body studies one is able to calculate this observable only for small systems. However, similarly to the other variables in this work, we are able to calculate it up to the order of $N=95$ spins. The Von Neumann entropy is an extensive quantity and in the following we will study the Von Neumann entropy per spin: $S\left(N\right)/N$. 
The mean-field entropy can be calculated by noting that the density matrix can be written in its Bloch sphere representation $\hat{\rho} = \frac{1}{2}\left(\mathbb{1} + \Vec{\epsilon}\cdot\Vec{\hat{\sigma}} \right)$. With $\Vec{\epsilon}$ the Bloch vector, which contains the magnetization in the $x$, $y$ and $z$-direction, and $\hat{\sigma}$ the Pauli matrices. The eigenvalues are given by $p = \left(1\pm \vert\Vec{\epsilon}\vert\right)/2$. These eigenvalues can be readily calculated from the steady-state mean-field equations \eqref{mfeq} and give access to the MF approximation of the Von Neumann entropy through \eqref{vn}, 
\begin{eqnarray}
    \frac{S_\text{MF}}{N} &=& -\frac{\left(1+ J\right)}{2}\ln\left(\frac{\left(1+ J\right)}{2}\right)
-\frac{\left(1- J\right)}{2}\ln\left(\frac{\left(1- J\right)}{2}\right),
    \nonumber\\&&
    \label{vnmf}
\end{eqnarray}
where $J^2=\langle \hat{S}^2\rangle=\text{Tr}[\hat{S}^2\hat{\rho}(t)]$ is the expectation value of the total spin length [c.f Eq.~(\ref{spinlength})] in the mean-field approximation.

The Von Neumann entropy solely depends on $J$ in Eq.~(\ref{vnmf}), illustrating the fact that states with maximum cooperation number, lying on the surface of the Bloch sphere, have minimum entropy. 
Instead, the entropy increases with decreasing spin length until the value $S_\text{MF}/N =\ln(2)$, which is indeed the maximum entropy of a qubit. 
In particular, we can express Eq.~(\ref{vnmf}) explicitly in terms of the steady-state values $\langle \hat{ \sigma}^x\rangle_{\rm ss}$, $\langle \hat{ \sigma}^y\rangle_{\rm ss}$ and $\langle \hat{ \sigma}^z\rangle_{\rm ss}$.
These results would be true independently of the model under consideration and even for the system dynamics, given the nature of the Gutzwiller-mean field ansatz for two-level systems.

\subsection{Bimodality coefficient} 
\label{sec:BC}
Using the permutational invariance present in this system, one is able to calculate results for a higher number of spins than usually feasible with other techniques. However, as noted before, finite-size effects are still present, hampering our ability to make a good estimate of the point of transition from the paramagnetic to the ferromagnetic phase using the order parameter. An indicator which is extremely suited for making a good estimate of this transition point is the bimodality coefficient, defined as
\begin{equation}
    B_c = \frac{m_2^2}{m_4},
\end{equation}
with $m_n$ being the $n$-th moment of an observable. The bimodality coefficient gives information on the bimodal nature of the operator used to calculate the moments. This bimodal nature indicates the presence of a ferromagnetic phase or a paramagnetic phase. A bimodal distribution for $\sum_i \sigma_i^x$, being the magnetization in the $x$-direction, indicates a ferromagnetic phase and typically has values close to $B_c = 1$. A paramagnetic phase, i.e. a unimodal distribution, is indicated by smaller values for $B_c$. A Gaussian distribution with zero mean has a value $B_c=1/3$ \cite{RotaNJP18,Chissom70}. 

Besides information on the nature of the phases at a specific parameter, the bimodality coefficient can also be used to indicate the transition point between the different phases. 
The curves for the bimodality coefficient for different system sizes intersect, providing an estimate of the critical point.
In finite-size systems, these intersection points coincide due to power-law dependence of correlations on the system size around the critical point.
In our case, since different number of spins correspond to different dimensions, this intersection point changes.
However, for sufficiently large systems this transition point should converge, indicating the phase transition.

We are interested in the presence of a ferromagnetic or paramagnetic phase in the $xy$-plane, and as such we study the emergence of ferromagnetic order in either the $x$ or $y$ direction. The second and fourth moments of $\hat{\sigma}_i^{x}$ and $\hat{\sigma}_i^{y}$ are readily calculated in the new basis, as they are expectation values of global operators. 

\begin{figure*}[!ht]
  \centering
    \includegraphics[width=\textwidth]{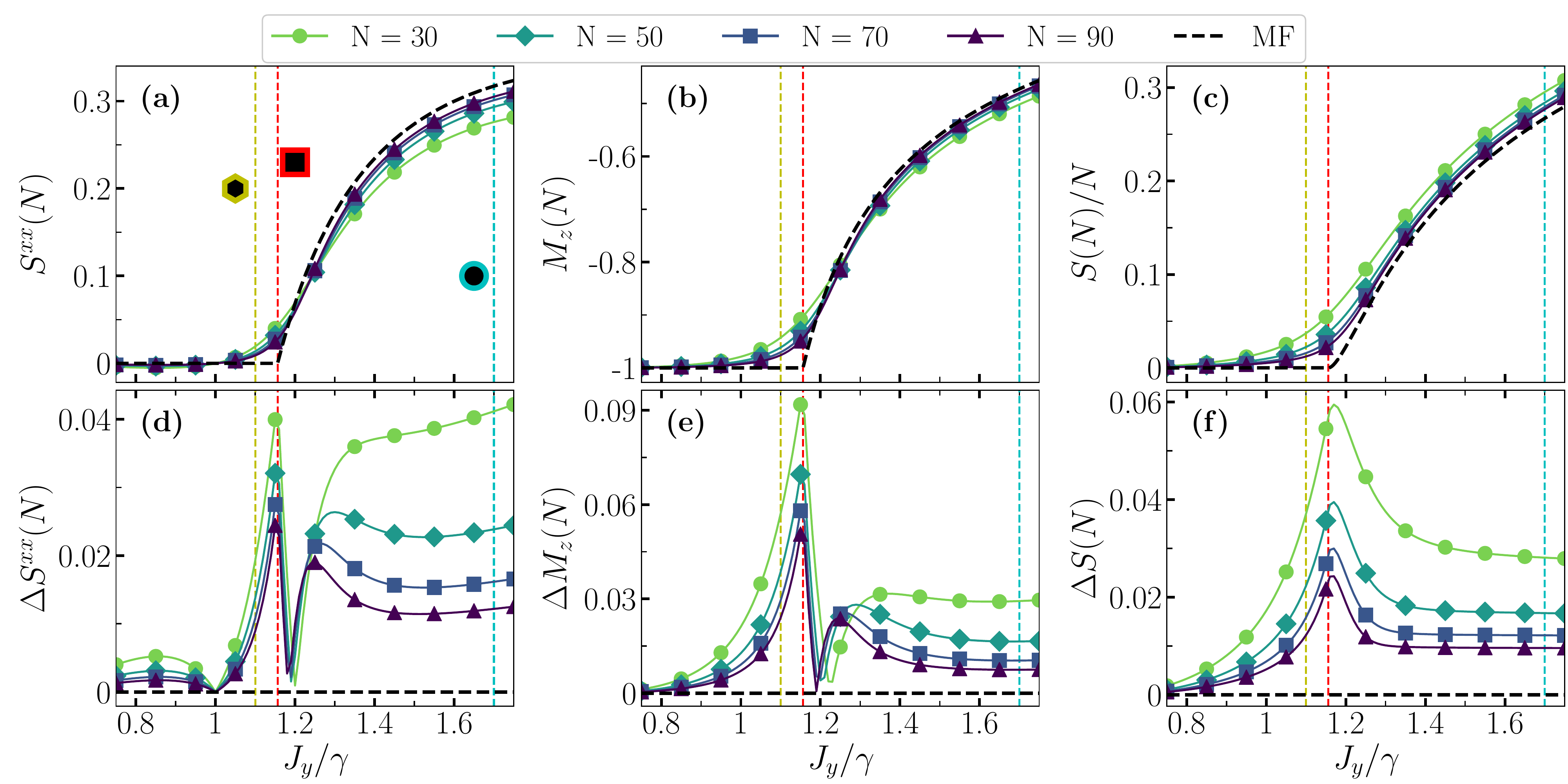}
      \caption{Study of the paramagnetic to ferromagnetic dissipative phase transition in the presence of only local dissipation for the system parameters specified in Fig. \ref{fig:MF_collective_vs_local} panel (b). The first row shows the steady-state spin structure factor in the $x$-direction [panel (a)], the $z$-magnetization [panel (b)], and the Von Neumann entropy per spin [panel (c)] as a function of $J_y$ for different system sizes ($N$ increases for darker curves). The markers are a guide for the eye, 100 points are calculated for each curve. The second row shows the absolute value of the difference between the variables in the corresponding upper panel and the mean-field value for $N\to\infty$. 
      (d) $\Delta S^{xx}(N)=\left(S^{xx}(N)-S^{xx}_\text{MF}(N)\right)/N$. 
      (e) $\Delta M_{z}(N)=\left(M_{z}(N)-M_{z \, \text{MF}}(N)\right)/N$. 
      (f) $\Delta S(N)=\left(S(N)-S_\text{MF}(N)\right)/N$. See Eq.~(\ref{mfn}) for details. In all panels, the black dashed curve represents the MF value.
      The dashed vertical lines refer to the points chosen in Fig.~\ref{fig:bcpd} and also studied for the system-size scaling in Fig.~\ref{fig:scaling}: the PM phase (yellow line, hexagon marker); the critical point (red line, square marker); the FM phase (cyan line, circle marker).} \label{fig:obs}
\end{figure*}

\begin{figure}
\centering
    \includegraphics[width=\linewidth]{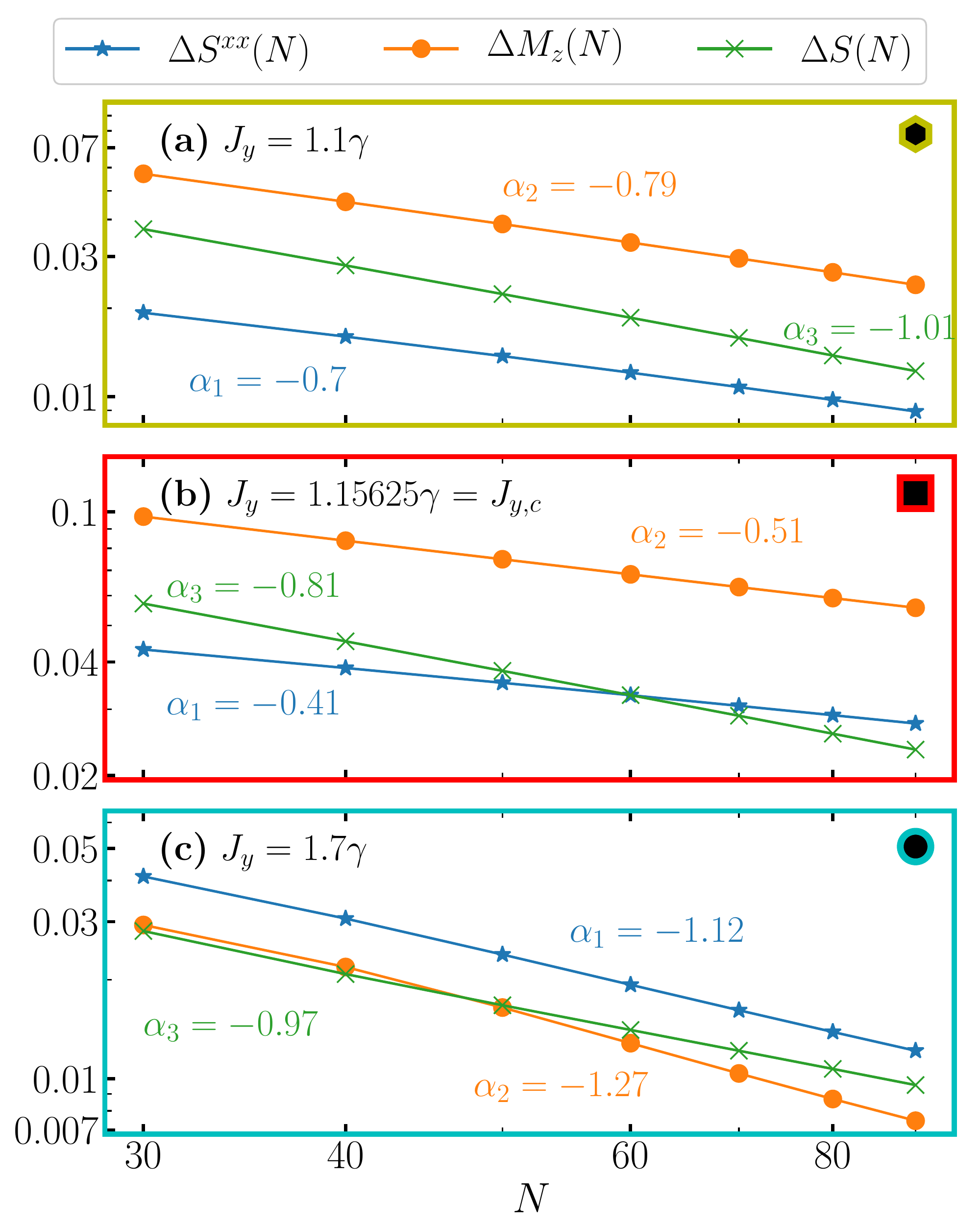}
      \caption{The panels show the finite size scalings of the quantities plotted in Fig.~\ref{fig:obs} for $J_y/\gamma = 1.1$ [panel (a)], $J_y/\gamma = J_{y,c}/\gamma$ [panel (b)], and $J_y/\gamma = 1.7$ [panel (c)]. We show the exponents $\alpha$ of a power law fit of the form $y = \beta N^{\alpha_i}$ next to the curves, for unknown coefficients $\beta$ and $\alpha_i$. The absolute difference of the spin-structure factor with respect to the MF prediction, for corresponding value of $N$, is marked by a blue line with stars and fit by $\alpha_1$. Similarly, in each panel the $z$-magnetization MF discrepancy is marked by an orange line with circles and exponent $\alpha_2$, while the Von Neumann entropy is marked by a green line with crosses, the exponent for the fit given $\alpha_3$. The markers in the top-right corner of each panel refer to the points in the phase diagram of Fig.~\ref{fig:bcpd}.} \label{fig:scaling}
\end{figure}

\subsection{Angular averaged susceptibility}
\label{sec:chiintro}
The paramagnetic-to-ferromagnetic phase transition is a second order one, and thus associated to a divergence of a response function.
The magnetic susceptibility informs us on the response of the system to a small external magnetic field and it is expected to diverge at the phase transition.
This is a consequence of the fluctuation-dissipation theorem, since fluctuations diverge at the critical point \cite{PathriaBOOK,Landau_BOOK_Statistical}. 
In a dissipative system, it is not always clear for which direction of the external perturbation the response should diverge.
Hence, in Ref.~\cite{RotaPRB17}, the concept of an \emph{angular averaged} magnetic susceptibility was introduced to study the \emph{XYZ}-model transition in a 2D lattice. 
If a small magnetic field of intensity $h$ is applied in the $xy$-plane as a probe,
\begin{equation}
    \hat{H}_B(h,\theta)= h\sum_i\left(\cos{\left(\theta\right)}\hat{\sigma}_i^x + \sin{\left(\theta\right)}\hat{\sigma}_i^y \right),
\end{equation}
it explicitly breaks the $\mathcal{Z}_2$-symmetry of the system.
By obtaining the perturbed steady state $\hat{\rho}(h, \theta)$ for $\hat{H}_{\rm ext}(h,\theta)=\hat{H} + \hat{H}_B(h,\theta)$, the resulting magnetization reads
\begin{equation}
    M_{\alpha}=\frac{1}{N} \sum_{j=1}^{N} \operatorname{Tr}\left[\hat{\rho}(h, \theta) \hat{\sigma}_{j}^{\alpha}\right], \qquad \alpha=x,\; y.
\end{equation}
Calling $h_x = h\cos{\left(\theta\right)}$ and $h_y = h\sin{\left(\theta\right)}$, the magnetic response in the linear regime is
\begin{equation}
    \Vec{M}\left(h,\theta\right) = \begin{pmatrix} \chi_{xx} & \chi_{xy} \\ \chi_{yx} & \chi_{yy} \\ \end{pmatrix}
    \begin{pmatrix} h\cos{\left(\theta\right)} \\ h\sin{\left(\theta\right)} \end{pmatrix},
\end{equation}
where the susceptibility tensor is defined as 
\begin{equation}
    \chi_{\alpha\beta} = \left.\frac{\partial M_\alpha}{\partial h_\beta}\right\vert_{h\rightarrow 0}.
\end{equation}
A scalar value can be obtained from this susceptibility tensor through angular averaging of the determinant, i.e.,
\begin{equation}
\label{eq:chi}
    \chi_{\rm av} = \frac{1}{2\pi}\int_0^{2\pi} \left.\frac{\partial\vert\Vec{M}\left(h, \theta\right)\vert}{\partial h}\right\vert_{h\rightarrow 0} d\theta.
\end{equation}

\section{Mean-field validity across the phase diagram}
\label{sec:validity}
Having introduced the main quantities and indicators which we will use to characterize the phase transition and the validity of the mean field, let us proceed to the numerical study of the model.

We use the permutational invariant quantum solver (PIQS) \cite{ShammahPRA98}, a module of QuTiP, the Quantum Toolbox in Python. This is an open-source computational library that leverages the flexibility of numerical and scientific Python libraries (NumPy and SciPy) and implements efficient numerical techniques by interfacing with the Intel Math Kernel Library (MKL). Performance is enhanced by using compiled scripts in Cython and by natively supporting cross-platform parallelization on clusters, with open multi-processing (Open MP) \cite{qutip1,qutip2}.
To obtain the steady-state density matrix, we will use the \verb|direct| method of the \verb|qutip.steadystate| solver, which is based on the lower-upper (LU) decomposition of the Liouvillian matrix to solve the equation $\LL \sss =0$.
The results are exact up to numerical tolerance (having set the absolute tolerance to $10^{-12}$)\footnote{The interested reader can find a series of notebooks dealing with similar systems in the section ``Permutational invariant Lindblad dynamics'' of the QuTiP project tutorials \href{http://qutip.org/tutorials}{http://qutip.org/tutorials}.}.

Based on the preliminary study of the Liouvillian gap, see Fig.~\ref{fig:gaps} and Sec.~\ref{sec:gap}, we can identify three main regions in the phase diagram of the \emph{XYZ} model: (i) Paramagnetic ($J_y\leq J_x$); (ii) Critical ($J_y \simeq J_x \simeq J_{y_c}$); (iii) High-anisotropy ($J_y > 2.3\gamma$), see discussion in Sec.~\ref{sec:HA}.
The paramagnetic one (i) seems to present a saturation of the Liouvillian gap and no antiferromagnetic phase for $J_y \leq 0$.
We may argue that this region can be safely approximated by a MF solution. We numerically tested this hypothesis, and found it to be correct (not shown). 

In the critical region (ii), a fundamental question is the determination of both the existence and the position of the critical point. 
Regardless of our ability to determine the point of transition, we are able to access the validity of the mean-field solutions through a finite size scaling. 
For almost-critical anisotropy, we will consider three domains: (1) the paramagnetic region before the transition, (2) the critical region according to MF prediction and (3) the ferromagnetic region.
Finally, we are interested in the properties of the high-anisotropy phase (iii). The MF does not predict a second phase transition to a paramagnetic phase. 
Nevertheless, several different methods \cite{LeePRL13,JinPRX16} have pointed out that this regime of parameters leads to a completely different behavior with respect to the standard ferromagnetic phase.

Note that in all the curves in this section which show the behavior of the system as a function of $J_y/\gamma$, the markers on the curves are a guide for the eye, and each curve is obtained from a simulation of a 100 points. We also computed more values of the system size $N$ than those shown in those figures.

In the following we choose, unless specified otherwise, $J_z = \gamma$, $J_x = 0.6\gamma$ and we vary $J_y$.

\subsection{Critical region}
\label{sec:critical}
In Fig.~\ref{fig:obs} we plot the spin structure factor [panel (a)], the $z$-magnetization [panel (b)], and the Von Neumann entropy [panel (c)] in the critical region $0.75<J_y/\gamma<1.75$ for different values of $N$, and we compare them to the results obtained via MF analysis (black dashed curve). 
According to Eq.~(\ref{eqinst}), we find 
\begin{equation}
S^{xx}_\text{MF}= (M_{x \, \text{MF}})^2 = 2 M_{z \, \text{MF}}\left(M_{z\, \text{MF}} + 1 \right)\frac{J_y - J_z}{J_x - J_y},
\end{equation}
 with 
\begin{equation}
M_{z \, \text{MF}} = -\frac{\gamma}{4}\frac{1}{\sqrt{\left(J_y - J_z\right)\left(J_z - J_x\right)}},
\end{equation}
where the mean field predicts a change between the PM and FM phases. Note that we use this point of transition as the definition of the critical point. The MF value of the Von Neumann entropy per spin is calculated using Eq.~\eqref{vnmf}.  
\\
All the three top panels of Fig.~\ref{fig:obs} show that the results of the full quantum simulations become closer to the MF prediction by increasing the number of sites.
Nevertheless, we notice that the results at the critical point are still in visible disagreement with respect to those obtained via MF analysis.

For our choice of parameters, $S^{xx}\left(N\right)>S^{yy}\left(N\right)$ and thus we study $S^{xx}\left(N\right)$. We identify a paramagnetic and a ferromagnetic phase in qualitative agreement with the mean-field calculations. Note that, as a result of finite-size effects, the transition from the paramagnet to the ferromagnet is smoothed, regardless of the observable being studied.
This makes it difficult to pinpoint the location of the phase transition using the spin structure factor as long as we are far from the thermodynamic limit.
Even more so as the region close to the transition is subject to sizeable fluctuations. We will return to the determination of the point of transition in subsection \ref{sec:bcvalidty}.

Normally, one expects the finite-size effects to disappear in the thermodynamic limit. 
To better quantify whether the exact quantum solutions would retrieve the mean-field results for $N\to \infty$, we study the absolute difference between the full quantum solution and the MF prediction for corresponding $N$,
\begin{subequations}
\label{mfn}
\begin{eqnarray}
\Delta S^{xx}(N)&=&|S^{xx}(N) -S_{\rm MF}^{xx}(N)|,\\
\Delta M_{z}(N)&=&|M_{z}(N) -M_{z \, {\rm MF}}(N)|,\\
\Delta S(N)&=&|S(N) -S_{\rm MF}(N)|,
\end{eqnarray}
\end{subequations}
for the steady-state spin structure factor, the $z$-magnetization, and the Von Neumann entropy, respectively. How these quantities fare as a function of $J_y$ is shown in panels (d-f) of Fig.~\ref{fig:obs}. The discrepancies are largest at the critical point (marked by a vertical red dashed line in each panel) and in general they tend to perform better in the anisotropic FM region, $J_y>J_z,J_x$ than in the PM region. We will better investigate the highly anisotropic region in Sec.~\ref{sec:HA}.
As a general trend, we can see that, as the system size is increased, the difference between the MF and the computed quantities from the quantum $\hat{\rho}_{\rm ss}$ becomes smaller. 
However, the three curves display different behaviors in their scaling properties.

In Fig.~\ref{fig:scaling} we show the finite-size scaling of the solution towards the MF, for the quantities of Eq.~(\ref{mfn}), for the three regions: (i) Paramagnetic, $J_y/\gamma = 1.1$ [panel (a)]; (ii) Critical, $J_y/\gamma = J_{y,c}/\gamma$ [panel (b)]; (iii) Ferromagnetic, $J_y/\gamma = 1.7$ [panel (c)].
We notice that all the results display a power-law behavior up to good approximation.
Thus, we perform a power-law fit of the form $y = \beta N^{\alpha_i}$ for unknowns coefficients $\beta$ and $\alpha_i$.
Clearly, $\alpha_i$ are negative for each observable, i.e., the mean-field solutions are in fact exact in the thermodynamic limit. 
However, different quantities in different regimes present different behaviors.
We notice that the ferromagnetic phase presents the highest convergence rate, the critical region being the slowest-converging one.
This is in accordance with the expected results, as the ferromagnetic region displays an ordered phase (low entropy) in which all the spin tends to be aligned, which can be better captured by a Gutzwiller ansatz.
Instead, at criticality, the system shows significant fluctuations around the MF results, which makes the convergence rate slower.

\subsection{Pinpointing the phase transition: Success of the bimodality coefficient and failure of the averaged susceptibility}
\label{sec:bcvalidty}

Having proved that the MF results recover the expected outcomes in the thermodynamic limit, we turn our attention now to the study of the critical point in finite size systems.
Indeed, in any experiment, one cannot access an infinite number of spins, but instead one has to infer the presence of criticality via finite-size scaling.
In this regard, we consider which quantity can better infer the existence of a phase transition in the thermodynamic limit.

\begin{figure*}[!h]
    \centering
    \includegraphics[width=\textwidth]{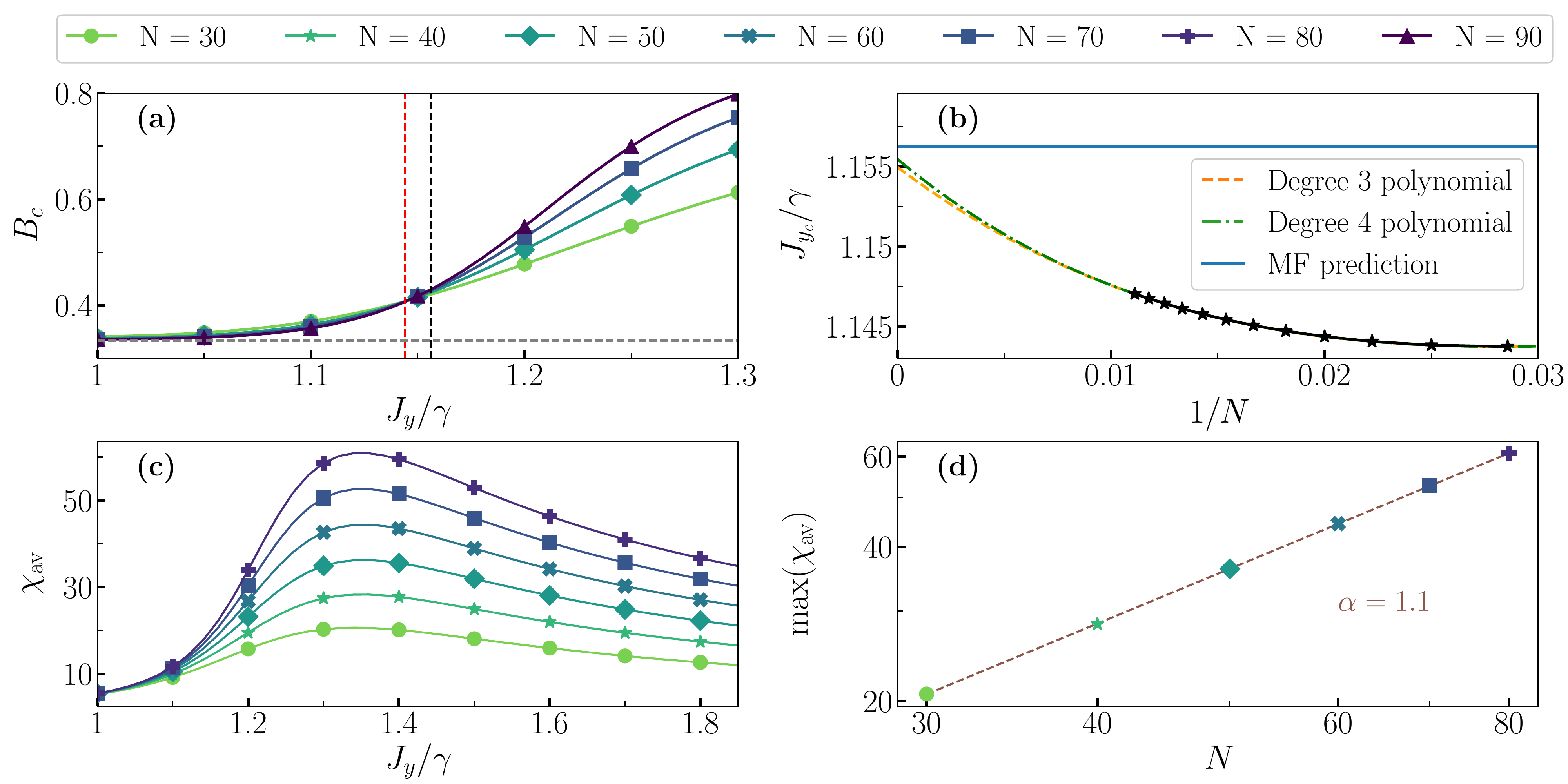} 
    \caption{Study of the location of the phase transition using bimodality coefficient (upper row) and the angular averaged susceptibility (lower row) for the system parameters specified in Fig. \ref{fig:MF_collective_vs_local} panel (b). (a) Bimodality coefficient in the $x$-direction. Where the critical point in the mean-field (black dashed line) is $J_{y,mf} = 1.15625\gamma$ and in the exact solution (red dashed line) $J_{y,e} = 1.144\gamma$, as determined by the intersection of the $N=50$ and $N=60$ curves. The (grey) horizontal dashed line indicates the value $1/3$, expected for the PM phase. (b) Point of transition as predicted by the intersection of the bimodality coefficient for systems with $N$ and $(N+5)$ spins (black full line with stars). The (blue) horizontal line indicates the mean-field prediction and the (orange) dashed and (green) dash-dotted curves respectively show a polynomial fit of degree three and four.
    The lower panels show a study of the angular averaged susceptibility for increasing system size $N$. (c) the  angular averaged susceptibility, $\chi_\text{av}$, is studied as a function of $J_y$. 
    (d) Scaling of the maximum of the angular averaged susceptibility as a function of the systems size $N$. The log-log fit extracts an exponent $\alpha=1.1$.  
    \label{fig:Bc}}
    
\end{figure*}

\begin{figure*}[!h]
  \centering
    \includegraphics[width=\textwidth]{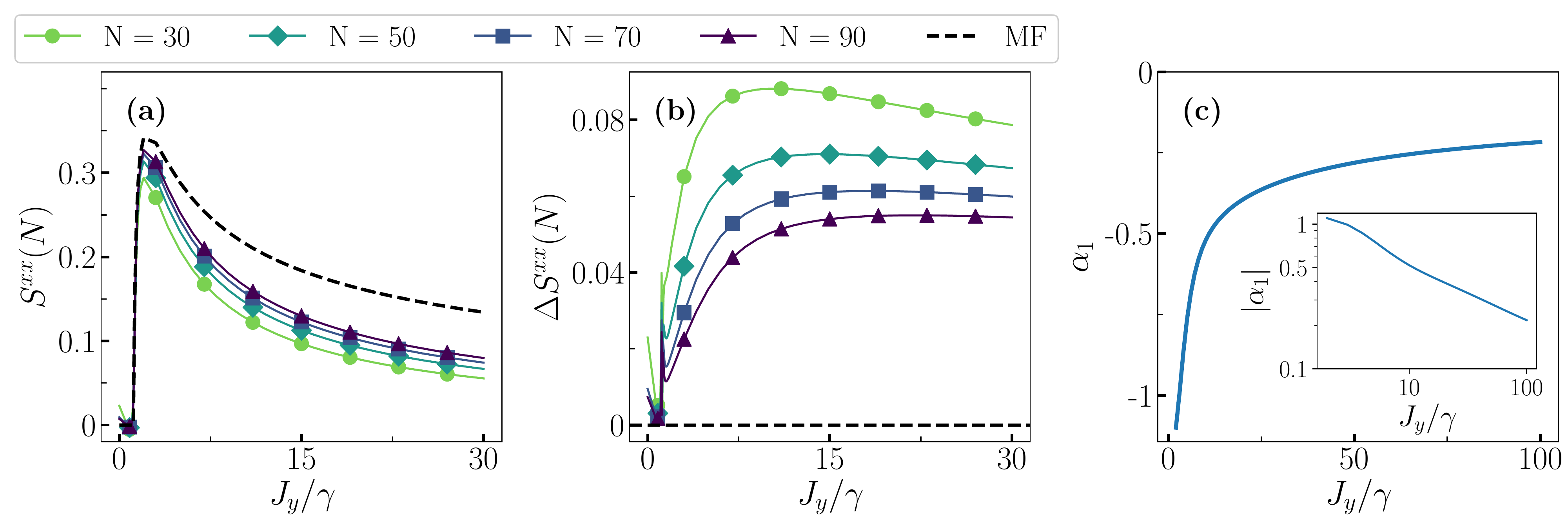}
      \caption{Study of the highly anisotropic ferromagnet and of the mean-field approximation validity, for local dissipation only. We set the system parameters as specified in Fig. \ref{fig:MF_collective_vs_local} panel (b) and study the spin structure factor as a function of $J_y$ for different system sizes (lighter to darker curves as $N$ increases). (a) Spin structure factor, $S^{xx}(N)$, calculated from the steady-state density matrix obtained from the Liouvillian in a fully-quantum picture. (b) Absolute difference between $S^{xx}(N)$ and the MF approximation for corresponding $N$.  (c) A power-law fit of the form $y = \beta N^{\alpha_1}$ is performed for $S^{xx}(N)$ for various points of $J_y$, using all the curves for different $N$ in panel (a), but up to the value $J_y/\gamma=100$. The inset highlights the variations in scaling with a log-log plot of $|\alpha_1|$. 
      } \label{fig:obslong}
\end{figure*}

\begin{figure*}
    \centering
    \includegraphics[width=\textwidth]{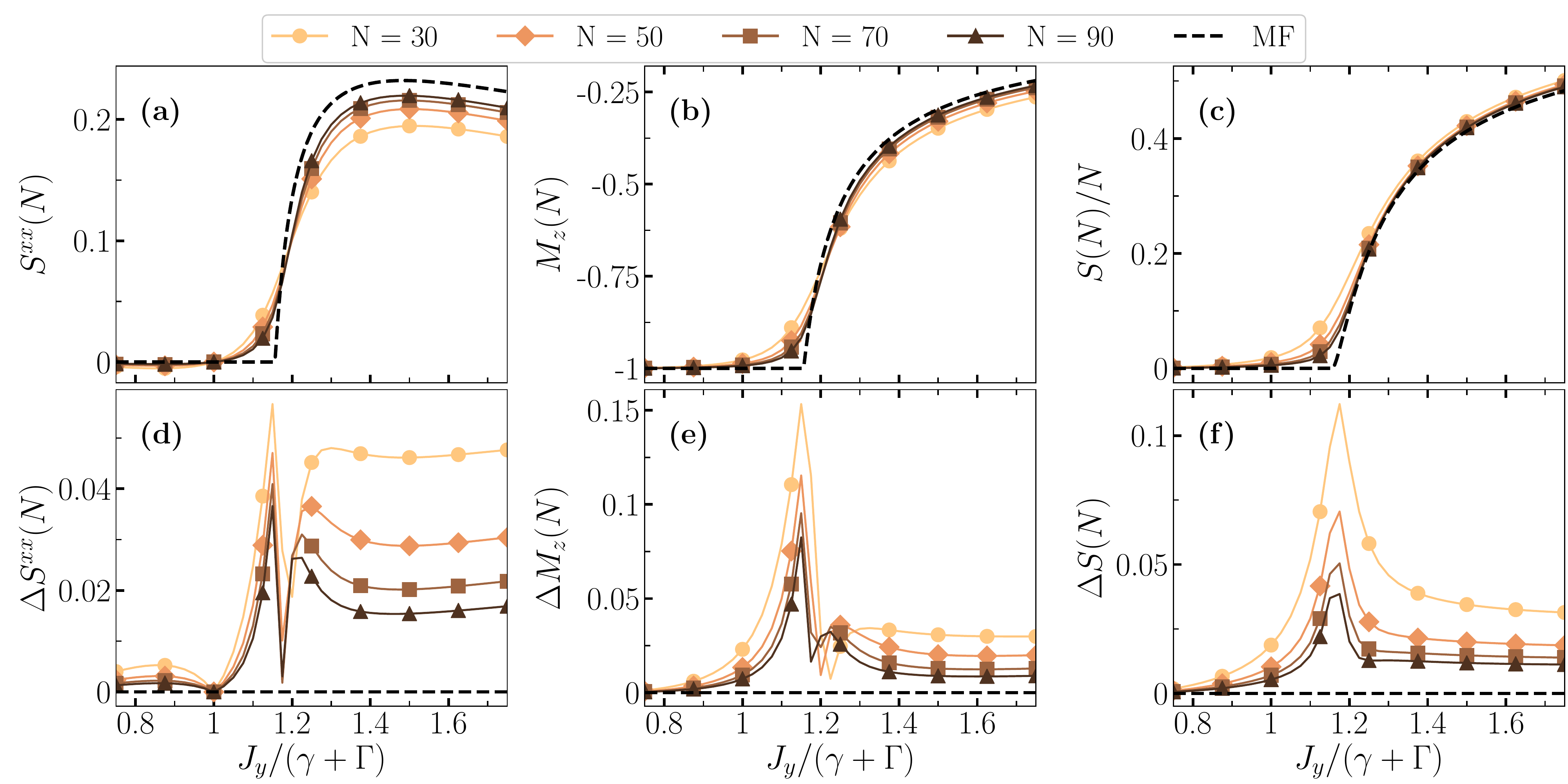}
    \caption{Study of the system in the presence of both local and collective dissipation near the paramagnetic to ferromagnetic dissipative phase transition for the system parameters specified in Fig. \ref{fig:MF_collective_vs_local} panel (a). The plots show the same quantities and parameter range for $J_y/(\gamma + \Gamma)$ as Fig.~\ref{fig:obs} (there $\Gamma=0$). (a) Spin structure factor, $S^{xx}(N)$. (b) $z$-magnetization, $M_{z}(N)$. (c) Von Neumann entropy per spin, $S(N)/N$. In all upper panels, the black dashed curve represents the MF value for $N\rightarrow\infty$. The lower panels highlight the difference with respect to the corresponding mean-field quantities for the same value $N$. The lower panels highlight the discrepancy with the mean field for fixed $N$, see Eq.~(\ref{mfn}).  
    (d) $\Delta S^{xx}(N)=\left(S^{xx}(N)-S^{xx}_\text{MF}(N)\right)/N$. (e) $\Delta M_{z}(N)=\left(M_{z}(N)-M_{z \, \text{MF}}(N)\right)/N$. (f) $\Delta S(N)=\left(S(N)-S_\text{MF}(N)\right)/N$.
    }
    \label{fig:my_label_coll}
\end{figure*}
\begin{figure}
    \centering
    \includegraphics[width=\linewidth]{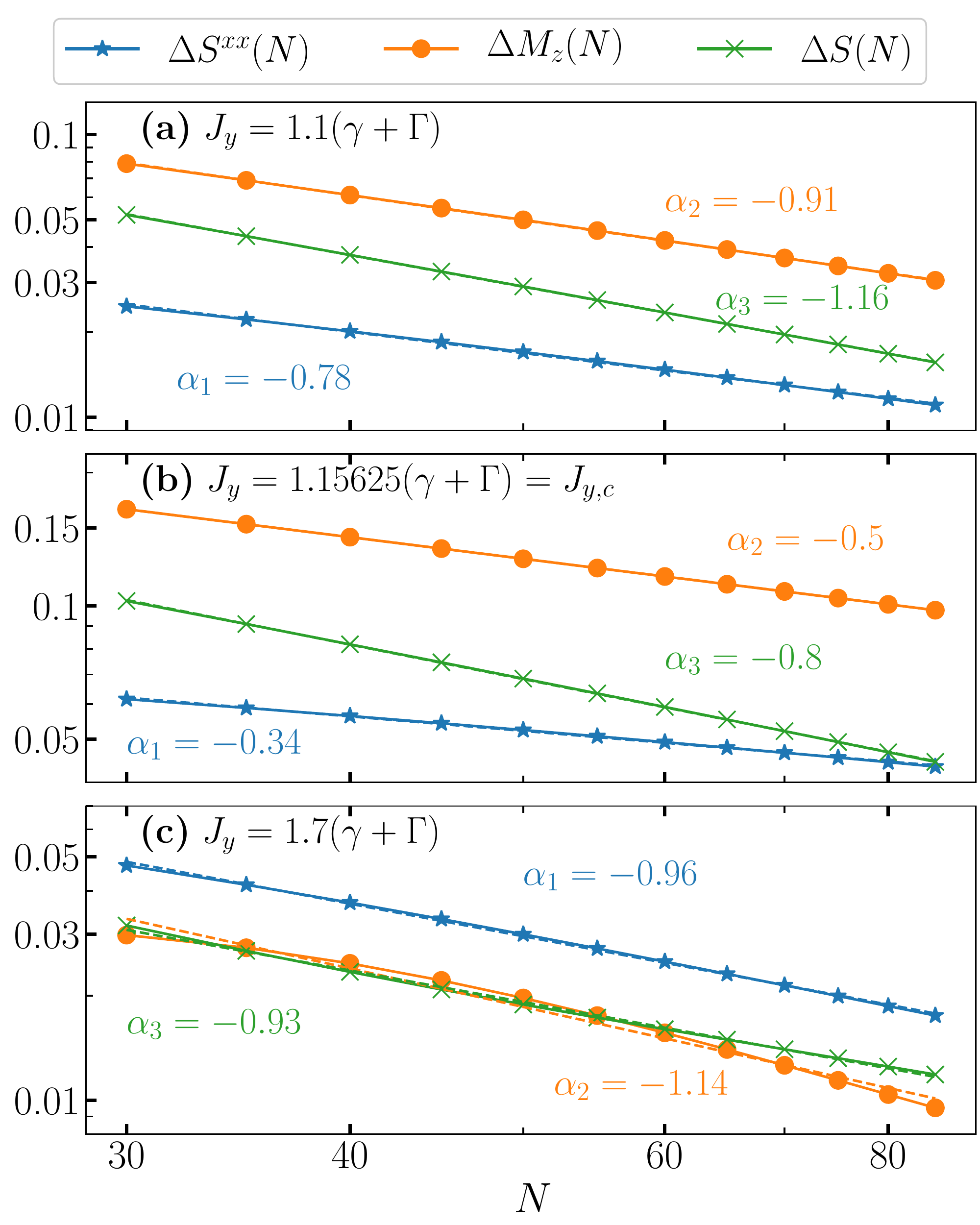}
    \caption{Study of the system-size scaling, extracted from the quantities plotted in Fig.~\ref{fig:my_label_coll}, in the presence of both local and collective dissipation across the paramagnetic to ferromagnetic dissipative phase transition. The same conventions as in Fig.~\ref{fig:scaling} are used to refer to the discrepancy between full-quantum simulation and MF prediction for the spin structure factor, the $z$-magnetization and the Von Neumann entropy. (a) We set $J_y=1.1\gamma$, (b) $J_y=J_{y,c}$ and (c) $J_y=1.7\gamma$.}
    \label{fig:my_label_coll_scaling}
\end{figure}

In panel (a) of Fig.~\ref{fig:Bc} we show results for $J_x = 0.6\gamma$. The vertical black dashed line shows the mean-field prediction for the position of the phase transition, the vertical red dashed line shows the position as predicted by the point of intersection of the bimodality coefficient between the curves $N=50$ and $N=60$. It is clear that finite-size effects impose a quantitative difference with the mean-field prediction for the location of the phase transition. 
Comparing the results for finite-size systems to those of the MF (Fig.~\ref{fig:bcpd}), the qualitative behavior is, however, in good agreement.
Moreover, the phases on either side of the transition coincide. On the left we see the values of the bimodality coefficient approaching $1/3$, indicating a unimodal, i.e., paramagnetic, region. And, on the right side, they approach $1$, indicating a bimodal region, i.e., a ferromagnetic one. 

One can wonder if there actually is a quantitative agreement in the thermodynamic limit and if not, how large the quantitative deviation from the mean-field value is. To gain a better idea of this we show the point of transition as predicted by the point of crossing of the bimodality coefficient curves for $N$ and $(N+5)$ as a function of $1/N$ in panel (b) of Fig.~\ref{fig:Bc}. As the system size increases, the point of transition moves towards the mean-field critical point. Even though we can simulate systems with a number of spins of the order of a hundred, we are still far away from the thermodynamic limit. To gain an estimate of the convergence in the thermodynamic limit we make a polynomial fit of third (orange dashed line) and fourth degree (green dash-dotted line). These results show us that in the thermodynamic limit the critical point is predicted with a reasonable, although not excellent, accuracy. 

In Fig.~\ref{fig:Bc}(c) and (d) we report on a study of the angular averaged susceptibility $\chi_{\rm av}$, as defined in Eq.~(\ref{eq:chi}). We find that this quantity \emph{is not a good predictor} of the position of the phase transition for finite number of spins $N$ in the all-to-all connected \emph{XYZ} spin model with local dissipation. 
Even if for small $N$ values the maximum of the susceptibility keeps shifting toward bigger $J_y/\gamma$ as $N$ increases, for bigger $N$ the peak is at a value $J_y\simeq 1.35 \gamma$ [Fig.~\ref{fig:Bc}(c)]. 
This value is different from that of the transition point predicted by the MF.
However, $\chi_{\rm av}$ becomes divergent for $N \to \infty$, as shown in panel (d). 
A log-log fit of the maximum extracts an exponent $\alpha=1.1$. 
We conclude that the angular averaged susceptibility, while signaling a divergence, is not associated to the one of the symmetry breaking. 
This is in stark contrast with lower dimensional cases \cite{RotaPRB17}.

\subsection{Highly anisotropic regime: Highly-entropic ferromagnet}
\label{sec:HA}
We now focus onto the study of the high-anisotropy regime.
We define it as the region of $J_y/\gamma$ where the phase is ferromagnetic but $S^{xx}$ decreases.
In our case, this corresponds to $J_y>2.3 \gamma$.
We verified that this point coincides exactly to that where the bimodality coefficient obtained via the MF solution starts to decrease.
In this regard, the high-anisotropy regime is the one where, by increasing $J_y$, the ferromagnetic phase peaks become less distinguished.

As already stated, this regime is particularly interesting.
Indeed, far from isotropy, the simultaneous creation of two spin excitations is energetically favorable. 
The Hamiltonian part tends to create correlations in the lattice while dissipation can act continuously to destroy them. 
The competition between the two actions creates very mixed and correlated states.
Indeed, the state remains very entropic even in the limit in which the Hamiltonian should dominate the dynamics.

Figure~\ref{fig:obslong} shows a detailed study of the steady-state spin structure factor in the $x$-direction.
We recall that in Fig.~\ref{fig:obs} we found that, for low anisotropy (i.e. $\vert J_x - J_y\vert$ small), the exact results converged quite well to the mean-field calculations, for the steady state spin structure factor as well for the other quantities. For large anisotropy, this appears no longer true, as illustrated by panel (a) up to $J_y/\gamma = 30$. In panel (b) we highlight the difference to the mean-field prediction, Eq.~(\ref{mfn}). A study on the scaling of the exponent,$S^{xx}(N)\propto N^{\alpha_1}$, is given in panel (c), for each point $J_y/\gamma$, up to $J_y/\gamma=100$, extracting the exponent for different values of $N$. 
Even though the scaling predicts a very slow convergence to the mean-field (e.g. $N^{-0.22}$ for $J_y/\gamma>60$) we derive a very different description of this regime. 
Since these exponents tend to zero for larger $J_y$ coupling, the MF prediction become less and less accurate the more we enter in the anisotropic regime.
The inset in \ref{fig:obslong}(c) provides a log-log scale of $|\alpha_1|$ versus $J_y/\gamma$ to even better illustrate the presence of different scaling regimes.
The plots of Fig.~\ref{fig:obs} and Fig.~\ref{fig:obslong} show that the correctness of the mean-field solutions depends on the parameter regime. More specifically: for low anisotropy it holds, and for larger anisotropy it does not.

We conclude that, even if there is not a second phase transition, in actual realization of the model the high-anisotropy regime can be seen as profoundly different from the low-anisotropy ferromagnet.
Not only does the order parameter in the MF become smaller and smaller, but the convergence of the full quantum solution towards the MF also becomes slower and slower. 
In this regard, the high-anisotropy region of the phase diagram seems to be inaccessible via experimental studies.

\subsection{Benchmark in the presence of local and collective dissipation}
\label{sec:loccol}
Finally, we consider the most general case in Eq.~(\ref{Eq:Lindblad}), for $\gamma\neq0$ and also $\Gamma\neq0$, i.e. we study the interplay of local and collective dissipation. The results of our numerical investigations are summarized in Fig.~\ref{fig:my_label_coll} and  Fig.~\ref{fig:my_label_coll_scaling}.  
The main observations are that the nature and position of the phase transition is not modified by the inclusion of collective dissipation, while some more refined qualitative features are affected, as also predicted by the MF solutions. 

Notably, the phase transition seems to become sharper, as highlighted both by the magnetization and spin structure behavior as a function of $J_y$ across the critical region, in panels (a) and (b) of Fig.~\ref{fig:my_label_coll}.
Similar features where observed when studying the Lipkin-Meshkov-Glick model with local and collective dissipation \cite{LeePRA14}.
The Von Neumann entropy, shown in panel (c), displays an excellent agreement with the MF prediction as the system size increases. Note that, similarly to Fig.~\ref{fig:obs}, the markers on the curves provide a guide for the eye, and 100 points are calculated for each curve as a function of $J_y$.
In the lower row of Fig.~\ref{fig:my_label_coll}, panels (d)-(f), we more precisely measure the difference from the MF result, showing that the highest discrepancies occur at the point of the phase transition and as the $J_y/\gamma$ normalized anisotropic coupling is increased. 

Moreover, in Fig.~\ref{fig:my_label_coll_scaling} we report the scaling of these quantities, as a function of $N$, in the PM region [panel (a)], at criticality [panel (b)], and in the FM region with moderate anisotropy with respect to the $|J_y-J_x|$ ratio [panel (c)]. 
Interestingly, panel (b) shows that at criticality, the same exponents as for the local dissipation case (see Fig.~\ref{fig:scaling}) for $\alpha_2$ ($z$-magnetization) and $\alpha_3$ (Von Neumann entropy per spin) are expected, with a slight discrepancy for $\alpha_3$ (spin structure factor). Similarly to the local-dissipation-only dynamics, in the FM anisotropic region, shown in panel (c), the system is well described by the MF even for low number of spins, as highlighted by $\Delta M_z(N)$, which decreases faster than a power-law behavior. Indeed, the magnetization absolute difference with respect to the MF displays a remarkable non-linear trend, that does not seem well captured by a linear fit in a log-log plot (a fit would produce $\alpha_2 =-1.14$, shown as a dashed orange curve).
This highlights the competition of processes governed by different scaling laws, hinting at the competition between local and collective dissipation even for remarkably large system sizes, $N\approx 100$.  

\section{Conclusions}
\label{sec:conclusions}
In this article we studied the steady-state properties of an all-to-all connected dissipative spin model and
tested the validity of the Gutzwiller mean-field approximation in capturing them. Specifically, we considered the benchmark model of the \emph{XYZ} anisotropic Heisenberg spin system, subject to both local and local-and-collective dissipation in the Lindblad form. 
This model is particularly interesting because it shows a second-order phase transition from a paramagnetic to a ferromagnetic phase. Moreover, for large anisotropy, this model presents a highly entropic regime which was debated to be a different phase according to cluster mean-field computations \cite{JinPRX16}. 

We simulated systems up to $N=95$ spins exploiting the permutational symmetry of the model \cite{ShammahPRA98}. We demonstrate that, in both cases, the mean field correctly captures the physics in the thermodynamic limit. However, the scaling in the low-anisotropy regime strongly differs from that in the high-anisotropy one: while in the former the agreement is also quantitative, in the latter the mean-field approximation fares worse. In this regard, we may advocate for the presence of strong correlations also in the all-to-all connected model.
Even if we find no signs of a second phase transition, we may still argue that the high-anisotropy ferromagnetic regime is physically different from the lower-anisotropy ferromagnet.  

Concerning more technical points, in absence of collective dissipation, we exploit the Liouvillian $\mathbb{PT}$-symmetry \cite{ProsenPRL12} to 
efficiently compute the spectral properties of the Liouvillian superoperator. In the presence of this weak symmetry the spectrum presents a second symmetry axis beyond the complex conjugation one. That, in turns, implies the existence of a state symmetric with respect to the steady state, and one associated to the first-excited eigenmatrix of the Liouvillian. The numerical computation of these two states is much easier than finding the real gap and steady state.
We thus introduced the antigap of $\mathbb{PT}$-symmetric Liouvillian systems, which is equivalent to the true Liouvillian gap, and thus marks criticality in open quantum systems \cite{KesslerPRA12,MingantiPRA18_Spectral}.  

The possibility to study a large range of spin system sizes allowed us to address the question of how to better characterize the emergence of criticality in finite-size systems. Our results indicate that the physics of systems out of equilibrium is more challenging to infer than one would naively expect, even in the best case scenario of all-to-all connected models, where dimensionality should induce a rapid decrease in correlations and fluctuations. 
Additionally, we have proven the resilience of the paramagnetic to ferromagnetic phase transition in the presence of both local and collective dissipation, finding that the presence of the two mechanisms does not change the nature of the phase transition.
In both cases, one observes a second order phase tranistion, and the onset of criticality is for the same parameters.
These indications are especially relevant to a broad variety of experimental platforms in which the dissipative phase transition can be studied, such as trapped ions, Rydberg atoms, superconducting circuits, and in solid state, especially with hybrid superconducting systems. More generally, these results provide a benchmark for the validity of mean-field approximations in understanding the experimental results obtained with noisy intermediate scale quantum simulators.

As a future outlook, we note that the interplay between local and collective dissipation beyond the all-to-all connected model demands further investigation with the adoption of both analytical and numerical approximate techniques. Exploiting other symmetries, such as translational invariance, it should be possible to further reduce the numerical resources for Liouvillian representation. Moreover, it will be interesting to investigate the system time evolution toward the steady state, as transient processes shall display even starker differences between mean-field or classical results and full quantum dynamics \cite{LeePRA14,OlmosPRA14,Schutz16,Defenu18,Pappalardi18,Gonzales19,Khasseh19}. Indeed, the present study focuses on the steady-state properties of the model, i.e., those which are permutationally invariant. Phenomena breaking this spatial symmetry, however, may arise in the dynamics towards the steady state.

\acknowledgements 
The authors acknowledge useful discussions with Alberto Biella, Riccardo Rota, and Wouter Verstraelen.
N.S. acknowledges hospitality by Marco Genoni and Matteo G.A. Paris in the Applied Quantum Mechanics group at the University of Milan, Italy. 
F.M. is supported by the FY2018 JSPS Postdoctoral Fellowship for Research
in Japan. F.N. acknowledges partial support from the MURI Center for
Dynamic Magneto-Optics via the Air Force Office of Scientific Research (AFOSR) award No. FA9550-14-1-0040, the
Army Research Office (ARO) under grant No. W911NF-18-
1-0358, the Asian Office of Aerospace Research and Development (AOARD) grant No. FA2386-18-1-4045, the Japan
Science and Technology Agency (JST) [through the Q-LEAP
program and CREST Grant No. JPMJCR1676], the Japan Society for the Promotion of Science
(JSPS) [through the JSPS-RFBR grant No. 17-52-50023 and
the JSPS-FWO grant No. VS.059.18N], the RIKEN-AIST
Challenge Research Fund, the FQXi and the NTT-PHI Labs. 
D.H. is supported by UAntwerpen/DOCPRO/34878. Part of the computational resources and services used in this work were provided by the VSC (Flemish Supercomputer Center), funded by the Research Foundation - Flanders (FWO) and the Flemish Government department EWI.

\newpage


%

\end{document}